\documentclass[11pt,a4paper]{article}
\pdfoutput=1
\usepackage{jheppub}
\usepackage{ifpdf}
\usepackage{afterpage}
\usepackage{amssymb}
\usepackage{amsmath}
\usepackage{graphicx}
\usepackage{subfig}
\usepackage{longtable}
\usepackage{verbatim}
\usepackage{amsfonts}
\usepackage{bbold}
\usepackage[table,usenames,dvipsnames]{xcolor}
\usepackage{titlesec}
\usepackage{cancel}
\usepackage{enumitem}
\usepackage{hyperref}
\usepackage[normalem]{ulem}
\usepackage{tabularx}

\usepackage{tikz}
\usetikzlibrary{trees}
\usetikzlibrary{decorations.pathmorphing}
\usetikzlibrary{decorations.markings}
\tikzset{
    photon/.style={decorate, decoration={snake}, draw=black},
    wino/.style={draw=redwine},
    electron/.style={draw=black, postaction={decorate},
        decoration={markings,mark=at position .55 with {\arrow[draw=black]{>}}}},
    scalar/.style={draw=black, dashed,postaction={decorate},
        decoration={markings,mark=at position .55 with {\arrow[draw=black]{>}}}},
    gluon/.style={decorate, draw=black,
        decoration={coil,amplitude=4pt, segment length=5pt}}
}

\newcommand{\bear}{\begin{array}}
\newcommand{\ear}{\end{array}}
\newcommand{\beq}{\begin{equation}}
\newcommand{\eeq}{\end{equation}}
\newcommand{\beqa}{\begin{eqnarray}}
\newcommand{\eeqa}{\end{eqnarray}}

\def\OMIT#1{{}}
\newcommand{\lsim}{\mathrel{\rlap{\lower4pt\hbox{\hskip1pt$\sim$}}
    \raise1pt\hbox{$<$}}}         
\newcommand{\gsim}{\mathrel{\rlap{\lower4pt\hbox{\hskip1pt$\sim$}}
    \raise1pt\hbox{$>$}}}         

\newcommand{\SM}{\mathrm{SM}}

\newcommand{\inter}{\mathrm{in}}
\newcommand{\exter}{\mathrm{ex}}
\newcommand{\ob}{\mathrm{ob}}
\newcommand{\fit}{\mathrm{fit}}

\newcommand{\eV}{\mathrm{eV}}

\newcommand{\Sec}[1]{Sec.~\ref{#1}}

\newcommand{\Fig}[1]{Fig.~\ref{#1}}
\newcommand{\Eq}[1]{Eq.~(\ref{#1})}

\newcommand{\ignore}[1]{}

\begin{document}
\title{\boldmath Constraints on photon mass and dark photon from the Jovian magnetic field}
\author[a]{Shi Yan,}
\author[a]{Lingfeng Li,}
\author[a,b]{JiJi Fan}
\affiliation[a]{Department of Physics, Brown University, Providence, RI, 02912, USA}
\affiliation[b]{Brown Theoretical Physics Center, Brown University, Providence, RI, 02912, USA}
\emailAdd{shi\_yan@brown.edu}
\emailAdd{lingfeng\_li@brown.edu}
\emailAdd{jiji\_fan@brown.edu}
\vspace*{1cm}

\abstract{The Jovian magnetic field, being the strongest and largest planetary one in the solar system, could offer us new insights into possible microscopic scale new physics, such as a non-zero mass of the Standard Model (SM) photon or a light dark photon kinetically mixing with the SM photon. We employ the immense data set from the latest Juno mission, which provides us unprecedented information about the magnetic field of the gas giant, together with a more rigorous statistical approach compared to the literature, to set strong constraints on the dark photon mass and kinetic mixing parameter, as well as the SM photon mass. The constraint on the dark photon parameters is independent of whether dark photon is (part of) dark matter or not, and serves as the most stringent one in a certain regime of the parameter space. }
\maketitle
\flushbottom

\section{Introduction}
\label{intro}

Jupiter, the largest planet in our solar system, with a stable and strong magnetic field, has attracted a lot of exploration and studies. Since the 1970s, 11 space missions have either flown by or orbited around Jupiter, each contributing to our understanding of this gas giant. Among these, the latest Juno mission stands out for its broad spatial coverage, ranging from approximately 3000 kilometers to Jupiter at its closest point to millions of kilometers at its furthest approaches. The wide span of its trajectory, combined with its extended mission duration, has allowed Juno to gather a wealth of data, offering invaluable insights into the planet's structure and magnetosphere~\cite{2017SSRv..213....5B}.

In addition to its significant contribution to planetary science, the Juno data could have some unexpected applications to probe and constrain new physics beyond the Standard Model (SM) of particle physics. One recent example is to use the {\it in situ} measurements on the relativistic electron fluxes to probe GeV-scale dark matter (DM) and long-lived dark mediator between the dark sector and our visible world~\cite{Li:2022wix}. There are a few other works applying the data of Jupiter (not the Juno data though) to constrain new physics, $e.g.$,~\cite{Leane:2021tjj,French:2022ccb,Liu:2023mll,Blanco:2023}. 

Other possible applications include constraining the photon mass and dark photon parameters with the Juno magnetic field data. In the SM, a photon is assumed to be massless, associated with an unbroken Abelian gauge group $U(1)_{\rm EM}$. However, there is no thorough theoretical argument against its possessing a (tiny) mass \cite{reece2019photon}. Another intriguing extension of the SM electromagnetic sector includes an additional dark $U(1)_d$ with a massive dark photon. This dark photon could kinetically mix with the SM photon, serving as a portal between the SM and the dark sector~\cite{Kobzarev:1966qya, Okun:1982xi, Galison:1983pa, Holdom:1985ag, Holdom:1986eq}. 
In both cases, the standard Maxwell equations need to be modified, and the predicted magnetic fields would change at length scales inversely correlated with the new mass scales introduced~\cite{Williams:1971ms,Bartlett:1988yy,Fischbach:1994ir}.
Jupiter's magnetic field, being the most substantial steady one accessible for {\it in situ} measurements, thus becomes a powerful source to constrain the massive photon and dark photon scenarios. Given that the data collected by Juno spans a spatial range comparable to the planet's size, it possesses the potential to unveil and constrain the inverse microscopic mass scale. In addition, the dark photon limit obtained in this manner is independent of its relic density. In other words, regardless of whether the dark photon constitutes part of the cold dark matter, the derived limit remains applicable.

In fact, using the Jovian magnetic field to constrain the photon mass is by no means a new idea. Back in 1975, Ref. \cite{davis1975limit} applied the Pioneer 10 data to set a photon mass bound, which is still being widely quoted today. The method's application to the Earth's magnetic field was even earlier~\cite{Goldhaber:1971mr}.
This analysis, however, faces several serious limitations. The data set from Pioneer 10 was relatively sparse, encompassing only a few hundred data points. Given its flyby trajectory, the mission did not offer a comprehensive exploration of Jupiter's magnetosphere, limiting the depth and breadth of the magnetic field data acquired. The constraint was derived from a overly simplified fit without quoting a statistical criteria. 

In the current Juno era, the Juno magnetometer (MAG) has provided an unprecedented data set, leading to the development of much more sophisticated Jovian magnetic field models. Notably, the Jovian Reference Models (JRM) such as JRM09~\cite{connerney2018new}, constructed using data from the initial periJoves (i.e., periapsises in orbit around Jupiter), and the more refined JRM33~\cite{connerney2022new}, which incorporates data from extended periJoves, offer detailed insights into the Jovian magnetic field. 
There already exists one study aiming to use the Juno information to constrain both the photon mass and the properties of dark photons~\cite{marocco2021dark}. 
Instead of using the full Juno MAG data, the study assumes that the Jovian magnetic field could be approximated by a dipole field. However, as those recent studies show, the Jovian magnetic field is much more complicated beyond a dipole field. Additionally, the study in~\cite{marocco2021dark} assumes the maximal deviation of the magnetic field to be about the size of the external field generated by the plasma current belt around Jupiter. These assumptions make the estimate simple and the corresponding constraint conservative, yet leaving room for improvements.

\begin{table}[h!]
    \centering
    \begin{tabularx}{\textwidth}{l|>{\arraybackslash}X}
    \hline
        Notation & Meaning \\
        \hline 
        $m_\gamma$ & photon mass \\
        $X_\mu$ & dark photon \\
        $m_X$  & dark photon mass\\ 
         $\varepsilon$ & kinetic mixing parameter \\
        \hline
        $r_J$ & Jupiter's radius \\
        \hline 
        $n$, $l$ &  indices for multipole expansions in spherical harmonics\\
        $i$ &  index for data points in the Juno MAG data\\
        $\alpha$ & index for Cartesian coordinates\\
        \hline
        $\vec{B}$, $\vec{J}$ & three-vectors of the magnetic field and electric current\\
        $R^{\rm in}$, $R^{\rm ex}$ & radial functions of the magnetic field from either internal or external sources in the massive photon scenario \\
        $\tilde{R}^{\rm in}$, $\tilde{R}^{\rm ex}$ & radial functions of the magnetic field from either internal or external sources in the dark photon scenario\\
        \hline
        $\sigma_{i,\alpha}$ & measurement uncertainty of the $\alpha$'th component of the magnetic field field at the $i^{\rm th}$ data point\\
        $\pmb{\Sigma}$ & covariance matrix of all the magnetic field components\\
        $\pmb{\psi}$ & collection of parameters describing new physics\\
        $\pmb{c}$ & collection of parameters describing the Jovian magnetic field\\
        $\pmb{\eta}$ & collection of all parameters including both $\pmb{\psi}$ and $\pmb{c}$ \\
        $\pmb{B}^\ob$ ($\pmb{B}^\fit)$ & collections of all observed (fitted) magnetic field components at all relevant data points (NOT a vector in the real space) \\
        \hline
        $p(\pmb{B}| \pmb{\psi})$ & marginalized likelihood function of a possible measurement $\pmb{B}$ after the magnetic field parameters $\pmb{c}$ are integrated out \\
        $\pi(\pmb\psi)$ & prior distribution of the model parameter set $\pmb{\psi}$\\
        $p(\pmb{\psi}|\pmb{B}^\ob)$ & posterior distribution of the model parameter set $\pmb{\psi}$ given the field measurement $\pmb{B}^\ob$\\
    \hline
    \end{tabularx}
    \caption{Important notations and their meanings.}
    \label{tab:notation}
\end{table}

In this paper, we will harness the original Juno MAG data to set constraints on the photon mass and dark photon parameters. We do not make the dipole approximation. Instead, we follow an approach similar to what is used to construct the sophisticated JRM models, with some further improvements. We implement a proper uncertainty estimation and include all the known contributions. We perform a complete statistical analysis to acquire the constraints. As a cross-check and a by-product, we demonstrate that our approach could reproduce the Jovian magnetic field in good agreement with the JRM models.  

The layout of the paper is as follows. In \Sec{math}, we present the modified Maxwell equations in the presence of a massive photon (\Sec{mass}) or a dark photon (\Sec{dark}), as well as the corresponding solutions of the magnetic field. In \Sec{var}, we discuss possible sources of the variance and how we estimate them. In \Sec{method}, we introduce the likelihood function and show how to use it to reconstruct the Jovian magnetic field. In \Sec{result}, we apply a Bayesian analysis with the Jeffreys prior to place constraints on the photon mass, as well as in the plane of dark photon mass and kinetic mixing parameter. We conclude in \Sec{conclusion}.

\section{Magnetic Field in the Presence of New Physics}
\label{math}

With new physics, Maxwell equations for electrodynamics could be modified. In this section, we review the corresponding modifications in two possible scenarios beyond the SM. In Sec.~\ref{mass}, we consider a massive SM photon, while in Sec.~\ref{dark}, we consider the scenario with a dark photon kinetically mixing with our photon.

\subsection{Massive Photon}
\label{mass}

In the SM, the photon is usually considered to be massless. However, there is not a theoretical argument proving that the photon mass has to be zero \cite{reece2019photon}. If the photon has a finite mass, the Lagrangian density for the electromagnetic field needs to be modified with an additional mass term
\begin{equation}
    \mathcal{L} \supset - \frac{1}{4} F_{\mu \nu} F^{\mu \nu} + e J^\mu A_\mu + \frac{m_{\gamma}^2}{2} A_\mu A^\mu~,
\end{equation}
where $F_{\mu \nu} = \partial_\mu A_\nu - \partial_\nu A_\mu$ is the field strength of the SM $U(1)_{\rm EM}$ with $A_\mu$ the four-vector potential, $e$ is the electric charge, $J_\mu$ is the current density, and $m_\gamma$ is the photon mass. 
The modified Ampère's law for the magnetic field $\Vec{B}$ in the presence of a non-zero photon mass can be obtained as 
\begin{equation}
    \nabla \times \vec{B} = -e  \Vec{J} - m_{\gamma}^2 \Vec{A}~,
    \label{mass_Amp}
\end{equation}
where $\Vec{B}, \Vec{A}, \Vec{J}$ are the three-component magnetic field, vector potential, and current, respectively.\footnote{It might be concerning that with a non-zero free electron density in the Jovian magnetosphere, the photon could pick up an effective plasma mass, see $e.g.$~\cite{Redondo:2008aa}. We first notice that such a plasma mass is a classic effect under the assumption of free electron gas. However, this assumption breaks down since the strong Jovian magnetic field is of $\mathcal{O}$(G), within which electrons are trapped along the field lines. In such cases the field is better described by the corresponding magnetohydrodynamics equations~\cite{marocco2021dark,DDRyutov_1997}, which further reduces to~\Eq{mass_Amp} for a static field.}
The three components of the internal field generated by the internal sources such as the dynamo current of Jupiter can then be expressed as
\begin{equation}
\left\{
\begin{array}{l}
    B_r^\inter = \sum_{n = 1}^\infty \sum_{l = 0}^n R_{1, n}^\inter(m_\gamma, r) [g^l_n \cos (l \phi) + h^l_n \sin (l \phi)] P^l_n (\cos \theta) \\
    B_\theta^\inter = - \sum_{n = 1}^\infty \sum_{l = 0}^n R_{2, n}^\inter(m_\gamma, r) [g^l_n \cos (l \phi) + h^l_n \sin (l \phi)] d_\theta P^l_n (\cos \theta) \\
    B_\phi^\inter = \frac{1}{\sin \theta} \sum_{n = 1}^\infty \sum_{l = 0}^n l R_{2, n}^\inter(m_\gamma, r) [g^l_n \sin (l \phi) - h^l_n \cos (l \phi)] P^l_n (\cos \theta) \\
\end{array}~,
\right.
\label{mass_in}
\end{equation}
where $r$, $\theta$ and $\phi$ are defined using the Jovian system III coordinate widely applied in geophysics~\cite{dessler1983physics}, which centers at the geometric center of Jupiter and rotates with the planet. The spherical and Cartesian coordinates to be discussed later are both defined with this coordinate system. $g_n^l$ and $h_n^l$ are the internal Schmidt coefficients, $P_n^l (\cos \theta)$'s are the Schmidt quasi-normalized associated Legendre functions of degree $n$ and order $l$, and $d_\theta$ is the derivative with respect to $\theta$. Plugging the equations above into \Eq{mass_Amp} and applying the magnetic Gauss's law $\nabla \cdot \Vec{B} = 0$, the radial functions are given by 
\begin{equation} \
\left\{
\begin{array}{l}
    R_{1, n}^\inter (m_\gamma, r) = \frac{m_\gamma^{n + 2} (n + 1)}{(2 n + 1)!!} \left[ k_{n + 1} \left( m_\gamma r \right) - k_{n - 1} \left( m_\gamma r \right) \right] \\
    R_{2, n}^\inter (m_\gamma, r) = \frac{m_\gamma^{n + 2}}{(2 n + 1)!!} \left[ k_{n + 1} \left( m_\gamma r \right) + \frac{n + 1}{n} k_{n - 1} \left( m_\gamma[] r \right) \right] \\
\end{array} ~,
\label{eq:Rin}
\right.
\end{equation}
where $k_n$ is the modified spherical Bessel function of the second kind with degree $n$. Note that the radial functions have the same dimension as the conventional multipole expansions ($r^{-n-2}$). In this work, $r$ is often expressed in the unit of Jupiter's radius ($r_J \approx 7.1 \times 10^4$~km), while products of the radial functions and the Schmidt coefficients, such as $R_{1, n}^\inter (m_\gamma, r) g^l_n$ or $R_{1, n}^\inter (m_\gamma, r) h^l_n$, are all in unit of Gauss. 

It is known that an external field generated by sources outside the planet itself must be taken into account for magnetic field modeling. The sources of such an external field vary, with the main contributor being the plasma belt current at around $r\gtrsim 7 r_J$~\cite{connerney2020jovian}. Inside the plasma belt region, the external field can be similarly expressed as
\begin{equation}
\left\{
\begin{array}{l}
    B_r^\exter = \sum_{n = 1}^\infty \sum_{l = 0}^n R_{1, n}^\exter(m_\gamma, r) [G^l_n \cos (l \phi) + H^l_n \sin (l \phi)] P^l_n (\cos \theta) \\
    B_\theta^\exter = - \sum_{n = 1}^\infty \sum_{l = 0}^n R_{2, n}^\exter(m_\gamma, r) [G^l_n \cos (l \phi) + H^l_n \sin (l \phi)] d_\theta P^l_n (\cos \theta) \\
    B_\phi^\exter = \frac{1}{\sin \theta} \sum_{n = 1}^\infty \sum_{l = 0}^n l R_{2, n}^\exter(m_\gamma, r) [G^l_n \cos (l \phi) + H^l_n \sin (l \phi)] P^l_n (\cos \theta) \\
\end{array} ~, 
\right.
\label{mass_ex}
\end{equation}
 where $G_n^l$ and $H_n^l$ are the external Schmidt coefficients and the corresponding radial functions are given as
\begin{equation}
\left\{
\begin{array}{l}
    R_{1, n}^\exter (m_\gamma, r) = \frac{-n}{m_\gamma^{n -1} (2 n + 1)!!} \left[ i_{n - 1} \left( m_\gamma r \right) - i_{n + 1} \left( m_\gamma r \right) \right] \\
    R_{2, n}^\exter (m_\gamma, r) = \frac{1}{m_\gamma^{n -1} (2 n + 1)!!} \left[ i_{n - 1} \left( m_\gamma r \right) + \frac{n}{n + 1} i_{n + 1} \left( m_\gamma r \right) \right] \\
\end{array}~,
\right.
\label{mass_R_ex} 
\end{equation}
where $i_n$ is the modified spherical Bessel function of the first kind with degree $n$. 

\subsection{Dark Photon}
\label{dark}

Now we turn to electrodynamics in the presence of a dark photon. The Lagrangian density of a dark photon kinetically mixing with the SM massless photon is given by \cite{holdom1986two}
\begin{equation}
    \mathcal{L} \supset - \frac{1}{4} \left( F_{\mu \nu} F^{\mu \nu} + X_{\mu \nu} X^{\mu \nu} \right) + \frac{\sin \kappa}{2} F_{\mu \nu} X^{\mu \nu} + e J^\mu A_\mu + \frac{m_X^2 \cos^2 \kappa}{2} X_\mu X^\mu~, 
\end{equation}
where $X_{\mu \nu} = \partial_\mu X_\nu - \partial_\nu X_\mu$ is the field strength of the dark $U(1)_d$, $X_\mu$ is the corresponding four-vector potential, $m_X$ is the mass of the dark photon, and $\kappa$ is the mixing angle. 
The kinetic mixing term $F_{\mu \nu} X^{\mu \nu}$ can be removed by diagonalization through $\Tilde{A}_\mu = A_\mu \cos \kappa$ and $\Tilde{X}_\mu = X_\mu - A_\mu \sin \kappa$. Thus, the Lagrangian density can be rewritten as
\begin{equation}
    \mathcal{L} \supset - \frac{1}{4} \left( \Tilde{F}_{\mu \nu} \Tilde{F}^{\mu \nu} + \Tilde{X}_{\mu \nu} \Tilde{X}^{\mu \nu} \right) + \Tilde{e} J^\mu \Tilde{A}_\mu + \frac{m_X^2}{2 (1 + \varepsilon^2)} \left( \Tilde{X}_\mu \Tilde{X}^\mu + 2 \varepsilon \Tilde{X}_\mu \Tilde{A}^\mu + \varepsilon^2 \Tilde{A}_\mu \Tilde{A}^\mu \right)~,
\end{equation}
where the mixing parameter is given by $\varepsilon \equiv \tan \kappa$, and $\tilde{e} = \sqrt{1 + \varepsilon^2}~e$. 

The dark photon-modified Ampère's law for static magnetic fields can be given as
\begin{equation}
\left\{
\begin{aligned}
    &\nabla \times \nabla \times \Vec{A} + \varepsilon^2 \frac{m_X^2}{1 + \varepsilon^2} \Vec{A} = - \Tilde{e} \Vec{J} - \varepsilon \frac{m_X^2}{1 + \varepsilon^2} \Vec{X} \\
    &\nabla \times \nabla \times \Vec{X} + \frac{m_X^2}{1 + \varepsilon^2} \Vec{X} = - \varepsilon \frac{m_X^2}{1 + \varepsilon^2} \Vec{A} \\
\end{aligned}~,
\right.
\end{equation}
where $\Vec{X}$ is the three-component vector potential for dark magnetic field.
To simplify these equations, we apply the following transformation $\Vec{\omega} = \Vec{A} - \varepsilon \Vec{X}$ and $\Vec{Q} = \Vec{A} + \Vec{X}/\varepsilon$. Then the equations above can be rewritten as
\begin{equation}
\left\{
\begin{aligned}
    &\nabla \times \nabla \times \Vec{\omega} = - \Tilde{e} \Vec{J} \\
    &\nabla \times \nabla \times \Vec{Q} + m_X^2 \Vec{Q} = - \Tilde{e} \Vec{J} \\
\end{aligned}~.
\right.
\label{massless_massive_equations}
\end{equation}
One can see that the equations for $\Vec{\omega}$ and $\Vec{Q}$ are similar to Ampère's laws for a massless photon and a massive photon with mass $m_X$, respectively, which have been discussed in \Sec{mass}.
The internal magnetic field is then
\begin{equation}
\left\{
\begin{array}{l}
    B_r^\inter = \sum_{n = 1}^\infty \sum_{l = 0}^n \tilde{R}_{1, n}^\inter(\varepsilon, m_X, r) [g^l_n \cos (l \phi) + h^l_n \sin (l \phi)] P^l_n (\cos \theta) \\
    B_\theta^\inter = - \sum_{n = 1}^\infty \sum_{l = 0}^n \tilde{R}_{2, n}^\inter(\varepsilon, m_X, r) [g^l_n \cos (l \phi) + h^l_n \sin (l \phi)] d_\theta P^l_n (\cos \theta) \\
    B_\phi^\inter = \frac{1}{\sin \theta} \sum_{n = 1}^\infty \sum_{l = 0}^n l \tilde{R}_{2, n}^\inter(\varepsilon, m_X, r) [g^l_n \sin (l \phi) - h^l_n \cos (l \phi)] P^l_n (\cos \theta) \\
\end{array}~,
\right.
\label{dark_in}
\end{equation}
where the radial functions are linear combinations of the massless and massive solutions as
\begin{equation}
\label{eq:darkphotonradialinternal}
\left\{
\begin{array}{l}
    \tilde{R}_{1, n}^\inter (\varepsilon, m_X, r) = \frac{n + 1}{1 + \varepsilon^2} \left( \frac{1}{r} \right)^{n + 2} + \frac{\varepsilon^2}{1 + \varepsilon^2} R_{1, n}^{\inter} (m_X, r)\\
    \tilde{R}_{2, n}^\inter (\varepsilon, m_X, r) = \frac{1}{1 + \varepsilon^2} \left( \frac{1}{r} \right)^{n + 2} + \frac{\varepsilon^2}{1 + \varepsilon^2} R_{2, n}^{\inter} (m_X, r)\\
\end{array}~,
\right.
\end{equation}
where $R_{1, n}^{\inter} (m_X, r)$ and $R_{2, n}^{\inter} (m_X, r)$ are the same as \Eq{eq:Rin} with $m_\gamma$ replaced by $m_X$. Here, the kinetic mixing $\varepsilon$ determines the relative importance between the two contributions in \Eq{eq:darkphotonradialinternal}. Larger $\varepsilon$ indicates that the solution deviates more from the SM. When deriving the equations above, there is a hidden assumption that the current generating the internal magnetic field is highly confined to the central region of Jupiter with $r\ll r_J$. It is explicitly shown through the $e^{-m_X r}$ factor in $R_{1, n}^{\inter} (m_X, r)$ and $R_{2, n}^{\inter} (m_X, r)$. If we assume that the currents are distributed elsewhere, like on the surface where $r=r_J$, by matching the boundary condition, the factor will become $e^{-m_X (r - r_J)}$ instead. In this case, the relative importance of the massive components increases, and the deviation from the SM will be more obvious, giving rise to a stronger constraint. From geophysics studies, the current distribution of Jupiter's dynamo is expected to be in a region of $r\lesssim (0.7-0.9)~r_J$~\cite{JONES2014148}. Considering the more complicated current distribution, which is beyond the scope of the paper, will only strengthen the bound we derive.

Similarly, the external magnetic field can be expressed as
\begin{equation}
\left\{
\begin{array}{l}
    B_r^\exter = \sum_{n = 1}^\infty \sum_{l = 0}^n \tilde{R}_{1, n}^\exter(\varepsilon, m_X, r) [G^l_n \cos (l \phi) + H^l_n \sin (l \phi)] P^l_n (\cos \theta) \\
    B_\theta^\exter = - \sum_{n = 1}^\infty \sum_{l = 0}^n \tilde{R}_{2, n}^\exter(\varepsilon, m_X, r) [G^l_n \cos (l \phi) + H^l_n \sin (l \phi)] d_\theta P^l_n (\cos \theta) \\
    B_\phi^\exter = \frac{1}{\sin \theta} \sum_{n = 1}^\infty \sum_{l = 0}^n l \tilde{R}_{2, n}^\exter(\varepsilon, m_X, r) [G^l_n \cos (l \phi) + H^l_n \sin (l \phi)] P^l_n (\cos \theta) \\
\end{array}~, 
\right.
\label{dark_ex}
\end{equation}
where the radial functions are given by
\begin{equation}
\left\{
\begin{array}{l}
    \tilde{R}_{1, n}^\exter (\varepsilon, m_X, r) = \frac{-n}{1 + \varepsilon^2} r^{n - 1} + \frac{\varepsilon^2}{1 + \varepsilon^2} \frac{r_0^{n - 1}}{R_{1, n}^{\exter} (m_X, r_0)} R_{1, n}^{\exter} (m_X, r)\\
    \tilde{R}_{2, n}^\exter (\varepsilon, m_X, r) = \frac{1}{1 + \varepsilon^2} r^{n - 1} + \frac{\varepsilon^2}{1 + \varepsilon^2} \frac{r_0^{n - 1}}{R_{2, n}^{\exter} (m_X, r_0)} R_{2, n}^{\exter} (m_X, r)\\
\end{array}~,
\right.
\end{equation}
where $r_0 = 7 r_J$. Here, we assume that all the plasma current that generates the external field is localized at 7~$r_J$. The effect of the external massive solution is highly limited. There will be no obvious difference in the fits even if we set $r_0$ to be infinity. 

\begin{figure}[h!]
    \centering
    \subfloat{\includegraphics[width=0.98\textwidth]{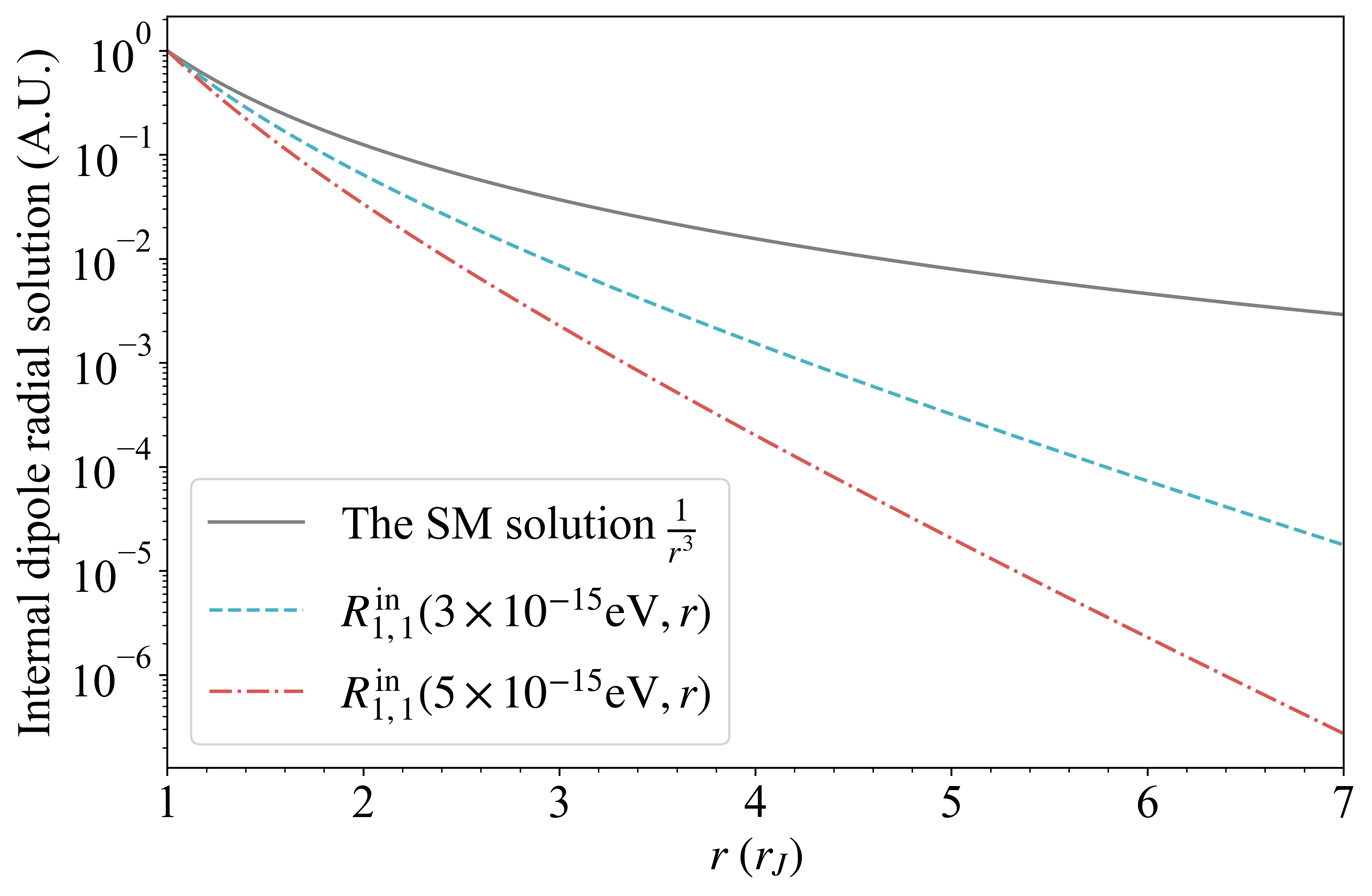}} \\
    \subfloat{\includegraphics[width=0.98\textwidth]{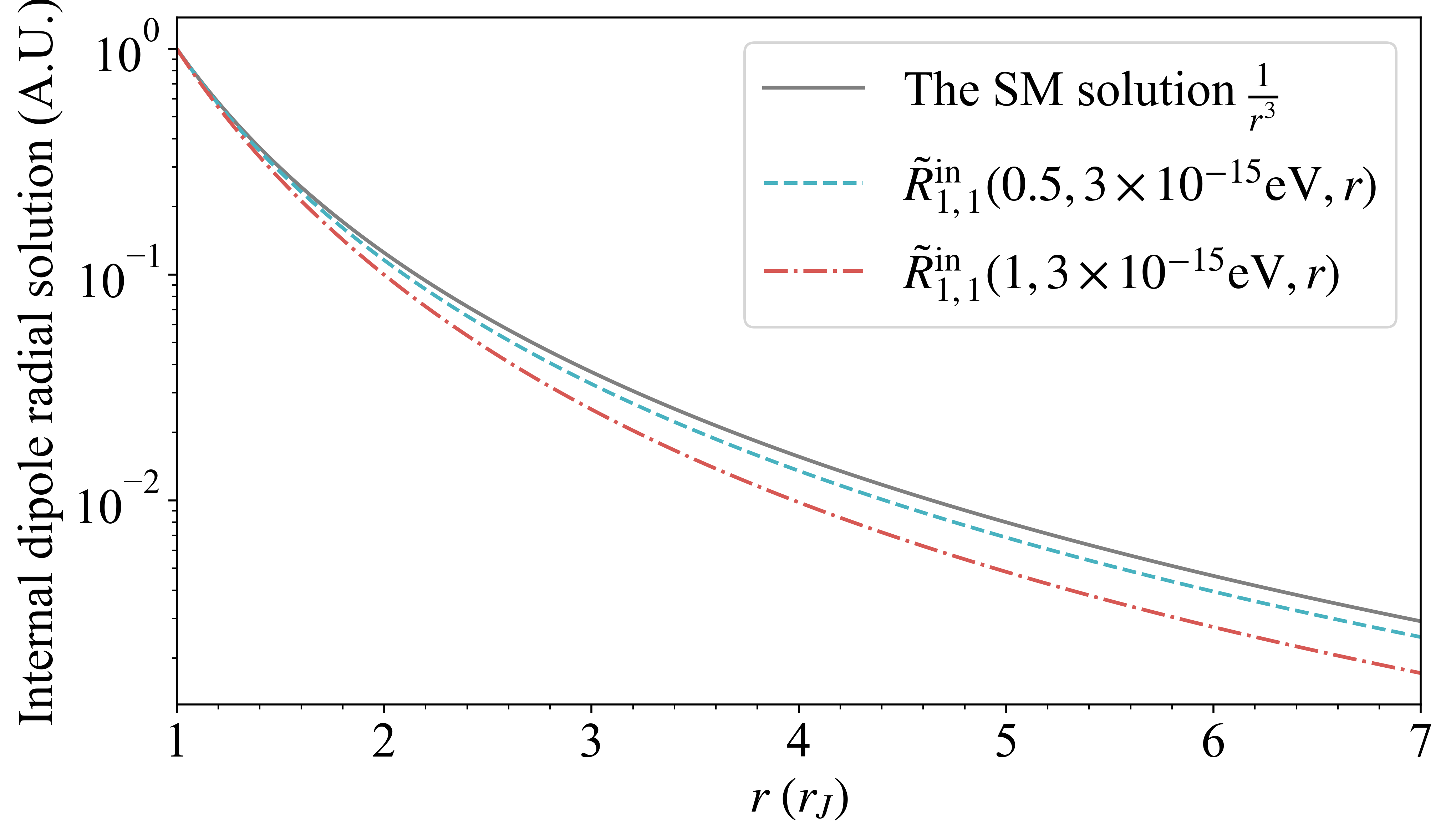}}
    \caption{Upper: the curves from top to bottom correspond to the dipole radial solution of the SM, $m_\gamma = 3 \times 10^{-15}$~eV and $m_\gamma = 5 \times 10^{-15}$~eV, respectively. Lower: the top curve corresponds to the solution of SM. The other two curves below correspond to the radial solution of the dark photon scenario with $m_X = 3 \times 10^{-15}$~eV, and $\varepsilon = 0.5$ and 1, respectively. The curves in both plots are normalized so that they match at $r =  r_J$. }
    \label{in_r}
\end{figure}

To show how the radial functions change in both the massive photon and dark photon scenarios, we use the magnetic dipole solution with $n = 1$ as an example and compare different radial solutions of the internal magnetic field in~\Fig{in_r}. As photon mass $m_\gamma$ increases, {\it e.g.}, from $3\times 10^{-15}$~eV to $5 \times 10^{-15}$~eV in the upper panel of~\Fig{in_r}, the radial solution, $R_{1,1}^{\rm in}$, deviates more from the SM solution, which indicates that the influence of the photon mass becomes more evident. Thus, there would be an upper limit on the photon mass since the Jupiter magnetic field is well modeled with the standard electrodynamics. When the mass of the dark photon $m_X$ is very large, {\it i.e.}, $m_X \gg r_J^{-1} \approx 2.8 \times 10^{-15}$~eV, the dark photon decouples from the low energy effective theory relevant for Jupiter. On the other hand, when it is very small, $m_X \ll r_J^{-1}$, its effect is also negligible. In both limits, the corresponding radial solution would be quite close to the SM solution. Thus for a given $\varepsilon$, the deviation of the radial solution would peak around $m_X \sim r_J^{-1} \approx 2.8 \times 10^{-15}$~eV. On the other hand, if we fix $m_X$ but change the kinetic mixing parameter $\varepsilon$ as shown in the lower panel of~\Fig{in_r}, the deviation increases as $\varepsilon$ increases, so that there will be an upper limit on $\varepsilon$ for each given $m_X$. 

\section{Sources of Variances}
\label{var}

We aim to conduct a first statistical analysis on constraining the photon mass and dark photon model using the Juno MAG data set. In particular, the data ranges from perijove 1 to perijove 40, with the perijove 2 data unavailable due to a spacecraft safe mode entry. The data set covers a time span of more than two years. To reduce the effect of an unaccounted external field, only the data taken within a distance of $r<7 r_J$ is adopted~\cite{connerney2020jovian}.

One crucial input for the analysis is the variances necessary for the likelihood functions. Variances we consider include $\sigma^2_{\mathrm{fluc}}$ from the short-term data fluctuation, $\sigma^2_{\mathrm{drift}}$ from the long-term drifting of the Jovian magnetic field, $\sigma^2_{\mathrm{dirc}}$ from the directional uncertainty of the Juno detector, and $\sigma^2_{\mathrm{res}}$ from the sensor resolution. These are either short-term variances with a time scale of $\lesssim{\cal O}({\rm min})$ or potentially much longer than the span of the Juno mission $\sim \mathcal{O}(\text{yr})$. We can then estimate their sizes for each data point locally or by taking the linear approximation with time. Other sources with intermediate time scales between $\mathcal{O}(\rm min)$ to $\mathcal{O}(\rm yr)$ include the magnetic field induced by the solar wind or complex plasma waves originating from the outer radiation belt. They still lack sound modeling in the literature and are not included in our analysis. With the variances we include, we manage to reconstruct the Jovian magnetic field in good agreement with the latest JRM models, as we will show in the next section.

The total variance $\sigma^2$ is then
\begin{equation}
    \sigma_{i, \alpha}^2 = \sigma^2_{\mathrm{fluc}, i, \alpha} + \sigma^2_{\mathrm{drift}, i, \alpha} + \sigma^2_{\mathrm{dirc}, i, \alpha} + \sigma^2_{\mathrm{res}, i, \alpha}~,
    \label{sig2}
\end{equation}
where the subscript $i$ is always used to indicate the $i$th data point while the subscript $\alpha$ indicates the $x, y, z$ spatial components. Below we will give more details for each contribution. 

\begin{figure}[h!]
	\centering
    \subfloat{\includegraphics[height=5 cm]{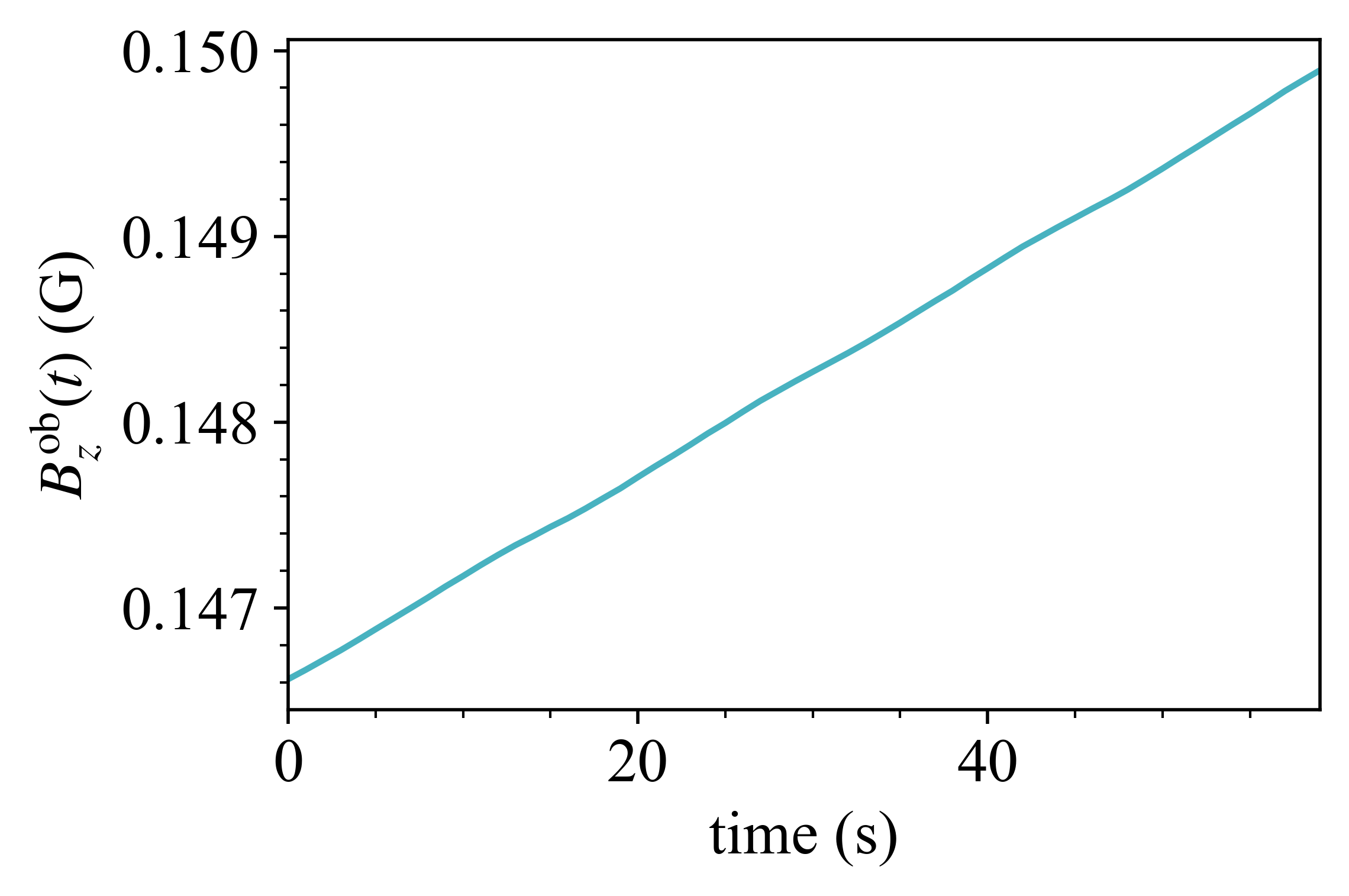}}
    \subfloat{\includegraphics[height=5 cm]{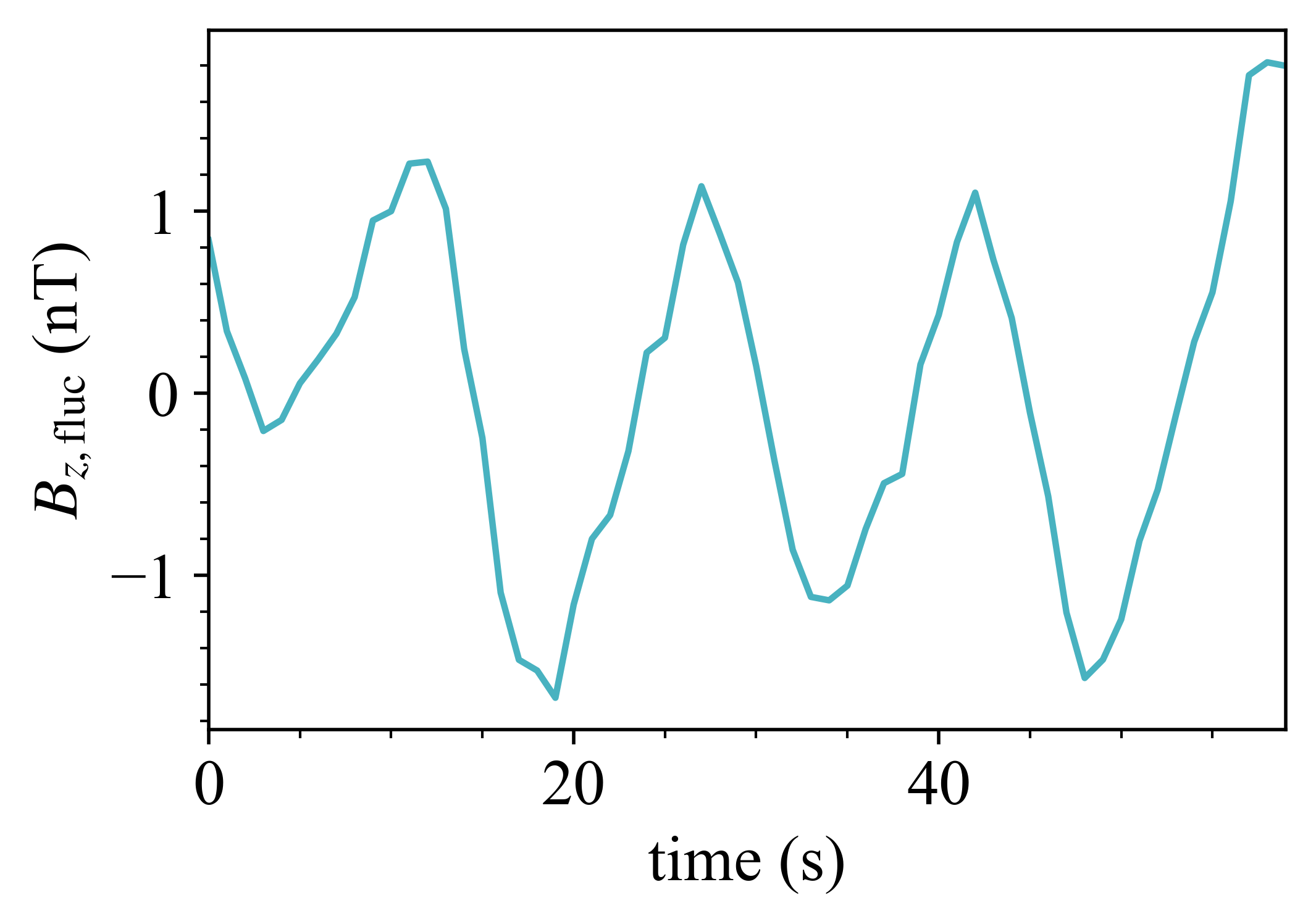}}\\
	\caption{Left: Juno MAG measurement of $B_z$ within one minute. Right: the residue $B_{z, \rm fluc}$ within one minute. Both plots are taken from a typical observation at $r = 3.6~r_J$ and obtained in day 240, 2016, 10:48-10:49. The raw data points are based on 1~second sampling.
     From the right panel, the obvious fluctuation with a period of $\simeq$ 15~seconds is induced by the eddy current. }
    \label{fluc}
\end{figure}

\noindent \textbf{Short-Term Fluctuations of Data} \quad  In \Fig{fluc}, we show that the raw data fluctuates over a time scale of one minute. The residual field shows a semi-periodic behavior after excluding the fast linear change due to the spacecraft's high velocity through the Jovian magnetic field. One major source of the short-term fluctuation is the eddy current induced by the Juno detector rotation~\cite{2020E&SS....701061K}. Juno spins with a period of 30 seconds. When its conductor cuts through the magnetic field, a secondary magnetic field modulating with a period of $\simeq$ 15 seconds is generated. Other sources could also contribute to the data fluctuation, but their natures still need to be understood. Thus, instead of describing all the uncertainties with time scales $\lesssim\mathcal{O}(\text{min})$ as an empirical function of Jovian coordinates, they will be extracted from the data directly.

Due to the short-term fluctuations, the data points of observed magnetic field $B_{i,\alpha}^{\ob}$ at time $t_i$ are averaged values around that time. Following the approach of the JRM09 model~\cite{connerney2018new}, the time interval $\delta t$ to be averaged over is chosen to be 60 s when $r_J < r <$ 4~$r_J$ and 120 s when 4~$r_J < r <$ 7~$r_J$
\footnote{The 4$R_J$ separation is empirical. Changing the separation would only affect the final limits mildly. For example, if we choose $\delta t = 120$ s everywhere, the estimated $\epsilon$ limit in the dark photon case will only be weaker by $\sim 25\%$.}
, so that the periodic fluctuations induced by Juno's spinning largely cancel out.
We start from the large raw data set containing $B_\alpha^{\rm ob}(t)$ measured with a much shorter sampling interval $\ll \delta t$. The averaged data points $B_{i,\alpha}^{\ob}$ is then related to $B_\alpha^{\rm ob}(t)$ as
\begin{equation}
    B_\alpha^{\rm \ob}(t) = B_{i, \alpha}^{\rm \ob} + B_{\mathrm{fluc}, i, \alpha} (t) + \frac{d B_{i, \alpha}}{d t} (t - t_i)~, |t-t_i|< \frac{\delta t}{2}~,
\end{equation}
where $B_{{\rm fluc},i,\alpha}(t)$ is the short-term fluctuation; $\frac{d B_{i, \alpha}}{d t} (t - t_i)$ is the linear variation of the measured field from Juno's movement along its orbit, which could be significant as shown in the left panel example of Fig.~\ref{fluc}. The corresponding fluctuating field $B_{\mathrm{fluc}, i, \alpha} (t)$, including possible higher-order terms, is plotted in the right panel of Fig.~\ref{fluc}. The semi-periodic behavior of $B_{\mathrm{fluc}, i, \alpha} (t)$ with a $\sim 15$~seconds period indicates that it is dominated by the eddy currents from Juno's spinning. However, apparent noises still contribute to this term with unclear origins. A complete component analysis of $B_{\mathrm{fluc}, i, \alpha} (t)$ is beyond the scope of this paper. The variance is computed from the fluctuations  $B_{\mathrm{fluc}, i, \alpha} (t)$ directly after filtering out the linear term due to the Juno motion as:
\begin{equation}
    \sigma^2_{\mathrm{fluc}, i, \alpha} = \overline{B_{\mathrm{fluc}, i, \alpha}^2}~.
\end{equation}

\noindent \textbf{Long-term Drifting of Jupiter's Internal Magnetic Field} \quad The Juno mission has spanned 5 to 6 years, and the magnetic field of Jupiter varies over this entire time window. The inferred change in Jupiter's magnetic field is consistent with the effects of secular variation, specifically that arising from the zonal drift of the internal structure of Jupiter. This term represents the long-term behavior of the field of order $\mathcal{O}~({\rm yr})$ and has been observed by multiple space missions~\cite{2019NatAs...3..730M}. This long-term drifting is mainly localized near the high-intensity spot near the equator~\cite{moore2017analysis, moore2018complex}, which is often referred to as the Great Blue Spot~(GBS).

With Schmidt coefficients of the differential zonal flux rotation (DFR) model~\cite{bloxham2022differential}, the change of Jovian magnetic field per year is obtained. The variance of each data point is estimated as
\begin{equation}
    \sigma^2_{\mathrm{drift}, i,\alpha} = \left[ D_{B, i, \alpha} (t_i - t_{\rm mid}) \right]^2~,
\end{equation}
where $D_{B, i, \alpha}$ is the long-term time variation of the magnetic field computed using the coefficients from the DFR model \cite{bloxham2022differential}. $t_{\rm mid}$ is a reference time which is the average of the starting and ending times of the current data set ($i.e.$, the time of the first data point in periJove 1 and the last data point in periJove 40). More specifically, it is in between periJove 19 and periJove 20. This choice of $t_{\rm mid}$ minimizes the variance, which will only be mildly modified if we vary $t_{\rm mid}$. Here we assume that the drifting speed of the field is approximately constant in the time span of our data set.

\noindent\textbf{Directional Uncertainty} \quad Juno's pointing attitude is determined by the Advanced Stellar Compass (ASC). The light from the stellar objects observed by the ASC will be affected by astronomical aberration caused by the moving spacecraft~\cite{connerney2017juno}. The aberrations effect shifts the apparent direction of the received light in the forward direction of motion. 

The magnitude of the aberration correction varies with the detector's velocity and many other factors. Conservatively, we choose to use the maximum aberration angle $\delta_{\mathrm{ab}} =  9.7\times 10^{-5}$ in the worst case scenario~\cite{connerney2017juno}. However, it is hard to determine the exact direction of such an uncertainty. To estimate the order of magnitude of this variance, we assume that the uncertainty is evenly distributed in all possible directions without correlation. This source turns out to be a subdominant contribution numerically, implying that the more complicated treatment will not change the overall variance significantly. To sum up, the variance from the directional uncertainty of Juno can be estimated as 
\begin{equation}
    \sigma^2_{\mathrm{dirc}, i, \alpha} = (\delta_{\mathrm{ab}} \; |\Vec{B}_i|)^2 / 3~,
\end{equation}
where $\Vec{B_i}$ represents the the $i$th observed magnetic field vector in the data set.  

\noindent\textbf{Detector Resolution (Intrinsic Digital Readout Uncertainty)} \quad The Juno sensor design covers a wide dynamic range with six instrument intervals increasing by a factor of four in successive steps. The resolution for each dynamic range is equal to half the step size for each interval~\cite{connerney2017juno}. The intervals (in the unit of G) and the corresponding resolutions, $\sigma_{\mathrm{res}}$, in the unit of nano Tesla (nT, 1 G = $10^5$ nT), are listed in Table~\ref{digital_noinse}.
\begin{table}[h!]
\centering
\begin{tabular}{c|c}
\hline
Dynamic Range (G) & $\sigma_{\mathrm{res}, i, \alpha}$ (nT) \\ \hline
$4.096 - 16.384$           & 25.0            \\
$1.024 - 4.096$            & 6.25            \\
$0.256 - 1.024$            & 1.56            \\
$0.064 - 0.256$           & 0.39           \\
$0.016 - 0.064$           & 0.19            \\
$\leq 0.016$           & 0.05            \\ \hline
\end{tabular}
\caption{Juno Fluxgate Magnetometer ranges and the corresponding readout resolutions~\cite{connerney2017juno}.}
\label{digital_noinse}
\end{table}

\begin{figure}[h!]
    \centering
    \subfloat{\includegraphics[width=0.5\textwidth]{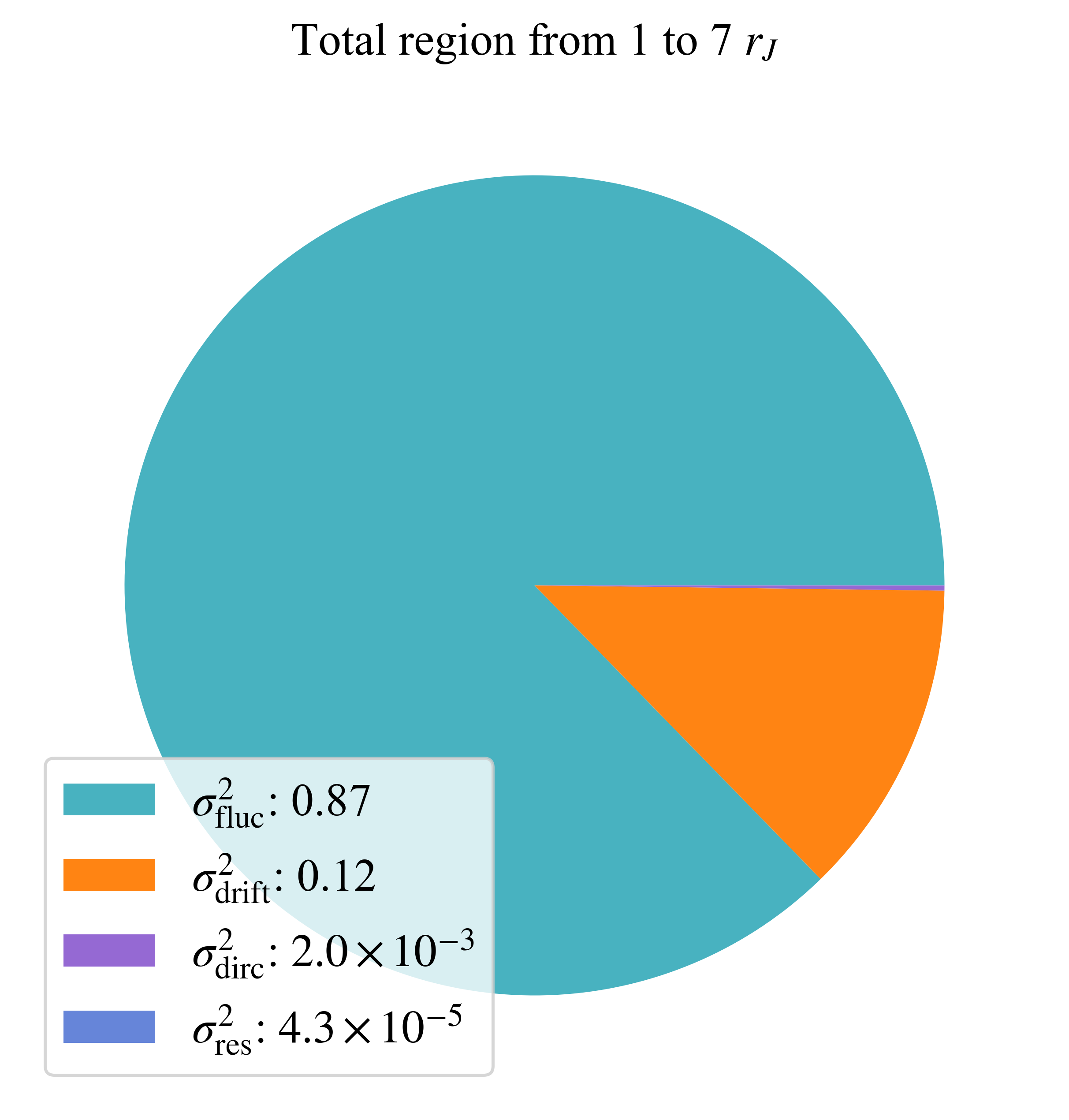}}
    \caption{Pie charts of components of the variances for the entire data set. 
    } 
    \label{pie}
\end{figure}

The relative importance of each variance for the data set considered is shown in \Fig{pie}. The proportion structures are similar in the innermost ($(1-2) r_J$) and outermost ($(6-7)r_J$) regions. The variances from the data fluctuation and the time variation of the Jovian magnetic field are the two dominant sources, while the remaining two sources, the directional uncertainty and the readout resolution, are negligible.  

\section{Magnetic Field Reconstruction}
\label{method}


Before presenting constraints on new physics, we will discuss how to reconstruct the Jovian magnetic field with the Juno data in this section. The entire model will be characterized by a set of parameters $\pmb{\eta} = \{ \pmb{\psi}, \pmb{c} \}$, in which $\pmb{\psi} = \{ m_\gamma \}$ for the massive SM photon scenario and $\pmb{\psi} = \{ m_X, \varepsilon \}$ for the dark photon scenario, are our parameters of interest, and the Schmidt coefficients $\pmb{c} = \{ G^0_1, G^1_1, H^0_1, \cdots, g^0_1, g^1_1, \cdots\}$ introduced in \Sec{math} describe the field configuration. The total number of the Schmidt coefficients is $M = n_{\rm max}^{\rm in} (n_{\rm max}^{\rm in} + 2) +  n_{\rm max}^{\rm ex} (n_{\rm max}^{\rm ex} + 2)$, where $n_{\rm max}^{\rm in}$ and $n_{\rm max}^{\rm ex}$ are the maximum degrees of the internal and external multipole expansions respectively. In this work, we choose $n_{\rm max}^{\rm in} = 18$ to be the same as the latest JRM model~\cite{connerney2020jovian}. More details of this choice could be found in Appendix~\ref{app:spectrum}. $n_{\rm max}^{\rm ex}$ is taken to be 5, with different choices leading to very small changes in the final results. Thus $M = 395$ in our study.

Given the complexity of the {\it in situ} measurements, getting the full correlations between magnetic field measurements and variances is much beyond the scope of this work. We assume that each observation $B^\ob_{i, \alpha}$, including those corresponding to different spatial components of the same data point, is independent of each other and Gaussian. Then we employ the following $\chi^2$ to describe the likelihood given by a fit:
\begin{equation}
    \chi^2(\pmb{\eta}) = \sum^{d}_{i = 1} \sum^3_{\alpha=1} \frac{(B_{i, \alpha}^\ob - B_{i, \alpha}^\fit(\pmb{\eta}))^2}{\sigma^2_{i, \alpha}}~,
\end{equation}
where $d$ is the total number of the data points from Juno used in this work, $B_{i, \alpha}^\ob$ is the value of the $i$th magnetic field data point's $\alpha$ component, while $B_{i, \alpha}^\fit$ is the prediction with the given parameter set $\pmb{\eta}$, and $\sigma^2_{i, \alpha}$ is the corresponding total variance defined in \Eq{sig2}. With this conventional definition, the likelihood function of a given parameter set (up to a global normalization factor) is $e^{-\chi^2}$. 


As a basic check of our method, we first try to find the magnetic field model $\pmb{c}$ that minimizes $\chi^2$ for a given set of model parameters $\pmb{\psi}$. Traditionally, such a coefficient set $\pmb{c}$ for the SM (equivalently $m_\gamma=0$ and no effect of dark photon due to either $\varepsilon=0$ or $m_X=0$) is considered to be the ``best-fit" Jovian magnetic field model describing the planet's interior dynamics, {\it e.g.}~\cite{connerney2017juno}. One important difference of the method is that in the geophysics literature, the residue of the $B$ field is minimized instead of $\chi^2$. 
To find the minimal $\chi^2$ for a given set of model parameters $\pmb{\psi}$, we can rewrite the solutions of the magnetic field given in \Sec{math} as a linear system in the Cartesian coordinate system as 
\begin{equation}
\Sigma^{-1} \pmb{B}^\fit = H(\pmb{\psi}) \pmb{c}~,
\label{linear}
\end{equation}
where $\Sigma$ is a $3 d \times 3 d$ matrix whose diagonal elements are the variances $\sigma^2_{i, \alpha}$, $\pmb{B}^\fit$ is a $3 d \times 1$ vector consisting of $B_{i, \alpha}^\fit$, and $H$ is a $3 d \times M$ matrix with elements given by the radial and angular functions in \Eq{mass_in} and \Eq{mass_ex} for the massive photon case, or \Eq{dark_in} and \Eq{dark_ex} for the dark photon case. Note that those equations in \Sec{math} are written in the spherical coordinate system while \Eq{linear} is written in the Cartesian system, which calls for a coordinate transformation. Then $\chi^2$ could be rewritten as 
\begin{equation}
    \chi^2(\pmb{\eta}) = \left[ \Sigma^{-1} \pmb{B}^\ob - H(\pmb{\psi}) \pmb{c} \right]^2~,
\end{equation}
where $\pmb{B}^\ob$ is a $3 d \times 1$ matrix consisting of $B_{i, \alpha}^\ob$.

The matrix $H$ can be expressed via the singular value decomposition (SVD) method~\cite{lanczos1996linear} as 
\begin{equation}
    H = U \Lambda V^T~,
\end{equation}
where $U$ and $V$ are $3 d \times 3 d$ and $M\times M$ unitary matrices, the $3 d \times M$ matrix $\Lambda$ is block diagonal, with $M$ diagonal values $\lambda_j$ are the singular values, whose squares are also eigenvalues of the matrix $HH^T$. Since the singular values $\lambda_j$ work as the system $H$'s eigenvalues, we will refer to $\lambda_j$ as eigenvalues without ambiguity for the rest of this paper. Each $\lambda_j$'s corresponding eigenvector $\pmb{v}_j$ is stored in the unitary matrix of $V$, which could be understood as a different basis of $\pmb{B}$ based on the data points instead of the multipoles. An eigenvector with a small eigenvalue represents a mode less constrained by the data since it leads to a smaller change in $\chi^2(\pmb{\eta})$. 
The Schmidt coefficients $\pmb{c}$ that minimizes the $\chi^2$ can thus be constructed by summing over the eigenvectors associated with a subset $k$ of the largest eigenvalues as
\begin{equation}
    \pmb{c} = \sum_{j = 1}^{k} \frac{\left( \Tilde{U}^T \Sigma^{-1} \pmb{B}^{\rm ob} \right)_j}{\lambda_j} \pmb{v}_j~,
\end{equation}
where $\Tilde{U}$ is an $3 d \times M$ matrix consisting of the first $M$ columns of the unitary matrix $U$.
Note that eigenvectors with small eigenvalues, which are less constrained, are omitted. In the case of the Juno MAG data, the least constrained eigenvectors often correspond to the $\pmb{B}$ field components that are only significant near Jupiter's south pole. This is because the inclination angle of the Juno orbits makes all the data points away from the region~\cite{connerney2017juno}.
If all eigenvectors are included, the magnetic field model starts to overfit, and the reconstructed field intensifies near the south pole. To reconstruct the $\pmb{B}$ field with sufficient precision while avoiding the overfitting problem, we follow~\cite{connerney2022new} and use only 300 out of 395 eigenvectors (i.e., $k = 300$) to construct $\pmb{c}$. More discussions can be found in Appendix~\ref{app:spectrum}. Removing the eigenvectors with the small eigenvalues leads to minor changes in the magnetic field reconstruction. One could then use $\pmb{c}$ to acquire the corresponding magnetic field. 

\begin{figure}[h!]
    \centering
    \subfloat{\includegraphics[width=\textwidth]{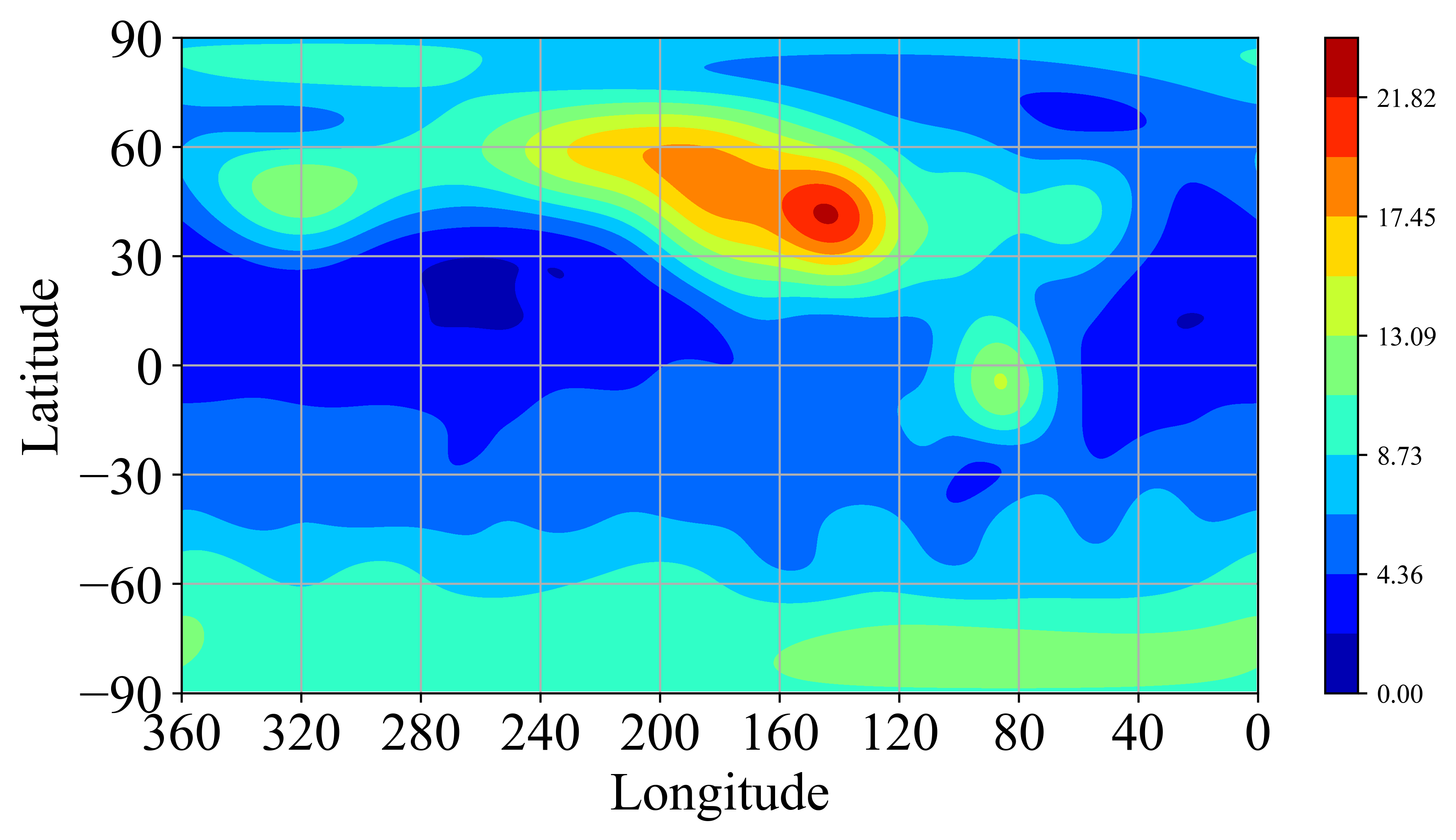}}\\
    \subfloat{\includegraphics[width=\textwidth]{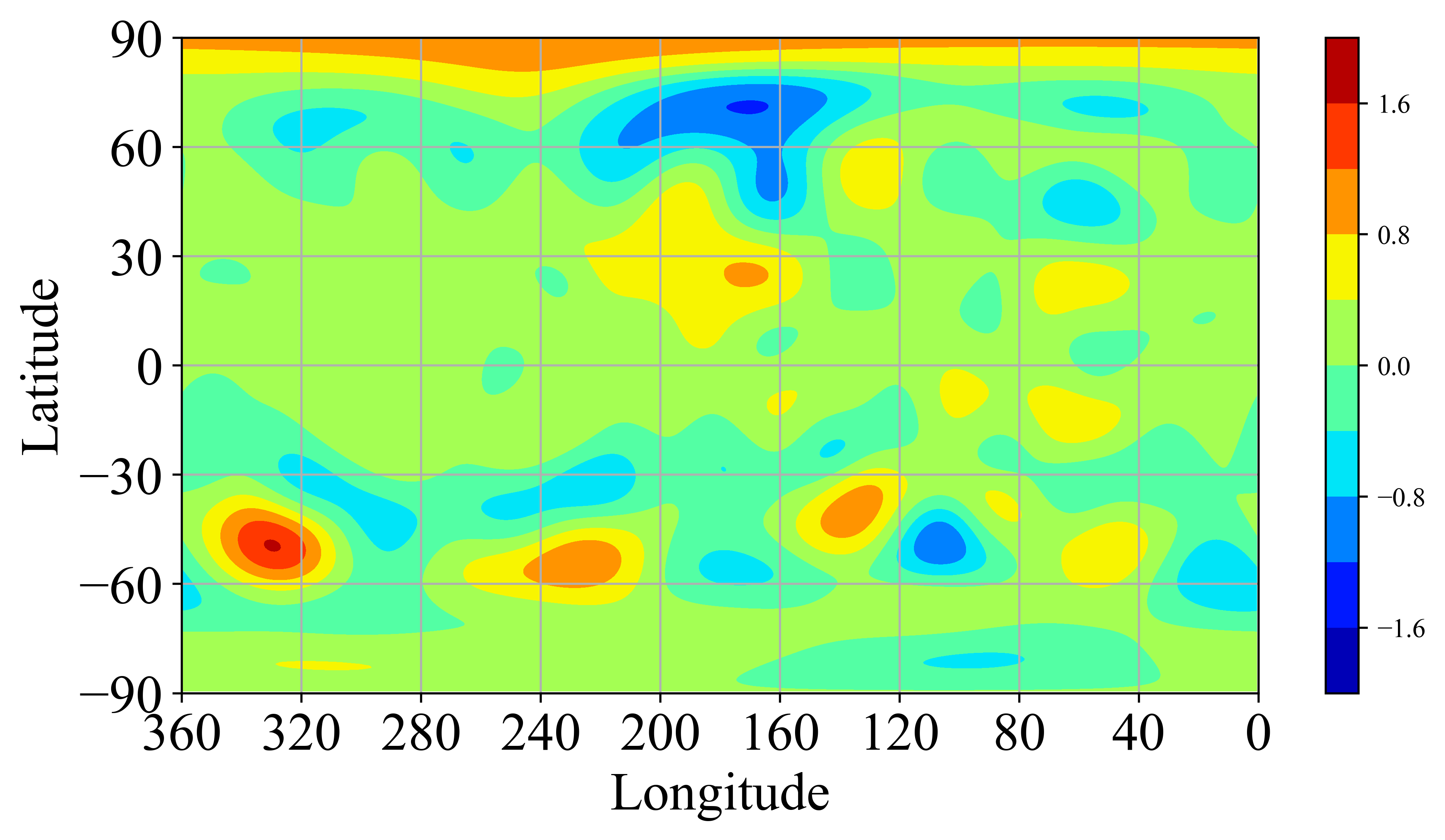}}
    \caption{Upper: contours of the field intensity $|\pmb{B}|$ (in unit of G) with $m_X = 10 ^{-18}$~eV and $\varepsilon = 1$. Lower:  contours of $|\pmb{B}| - |\pmb{B}_\SM|$ (in unit of nT$=10^{-5}$ G) with $\pmb{B}_\SM$ being the best-fit SM magnetic field. Both contours are plotted on the dynamically flattened (1/15.4) surface of Jupiter via the rectangular latitude-longitude projection.}
    \label{r_map}
\end{figure}

In \Fig{r_map}, we show examples of the reconstructed magnetic field that minimizes $\chi^2$ for a given set of new physics parameters. The upper panel of \Fig{r_map} shows the contours of the field intensity on the surface of Jupiter\footnote{Since Jupiter is not completely spherical, we take the dynamical flattening into account, which is $\sim 1/15.4$.} from $\pmb{c}$ minimizing $\chi^2$ when $m_X = 10 ^{-18}$~eV$\approx (3 \times 10^3 \; r_J)^{-1}$ and $\varepsilon = 1$. The lower panel of \Fig{r_map} presents the contours of the difference between the field in the upper panel and the one based on the SM. Since $m_X = 10 ^{-18}$~eV is small, as we have discussed at the end of \Sec{math}, the effect from the dark photon is tiny and the map is very similar to the one in the SM case \cite{connerney2022new}. In practice, we find that the marginalized likelihoods after integrating out $\pmb{c}$ are largely governed by $\chi^2_{\rm min}(\pmb{\psi})$, or equivalently the maximal likelihoods. Consequently, the behaviour of $\chi^2_{\rm min}(\pmb{\psi})$ sheds light on where the new physics could be constrained with the Juno MAG data.

\begin{figure}[h!]
	\centering
	\subfloat{\includegraphics[width = 0.45\textwidth]{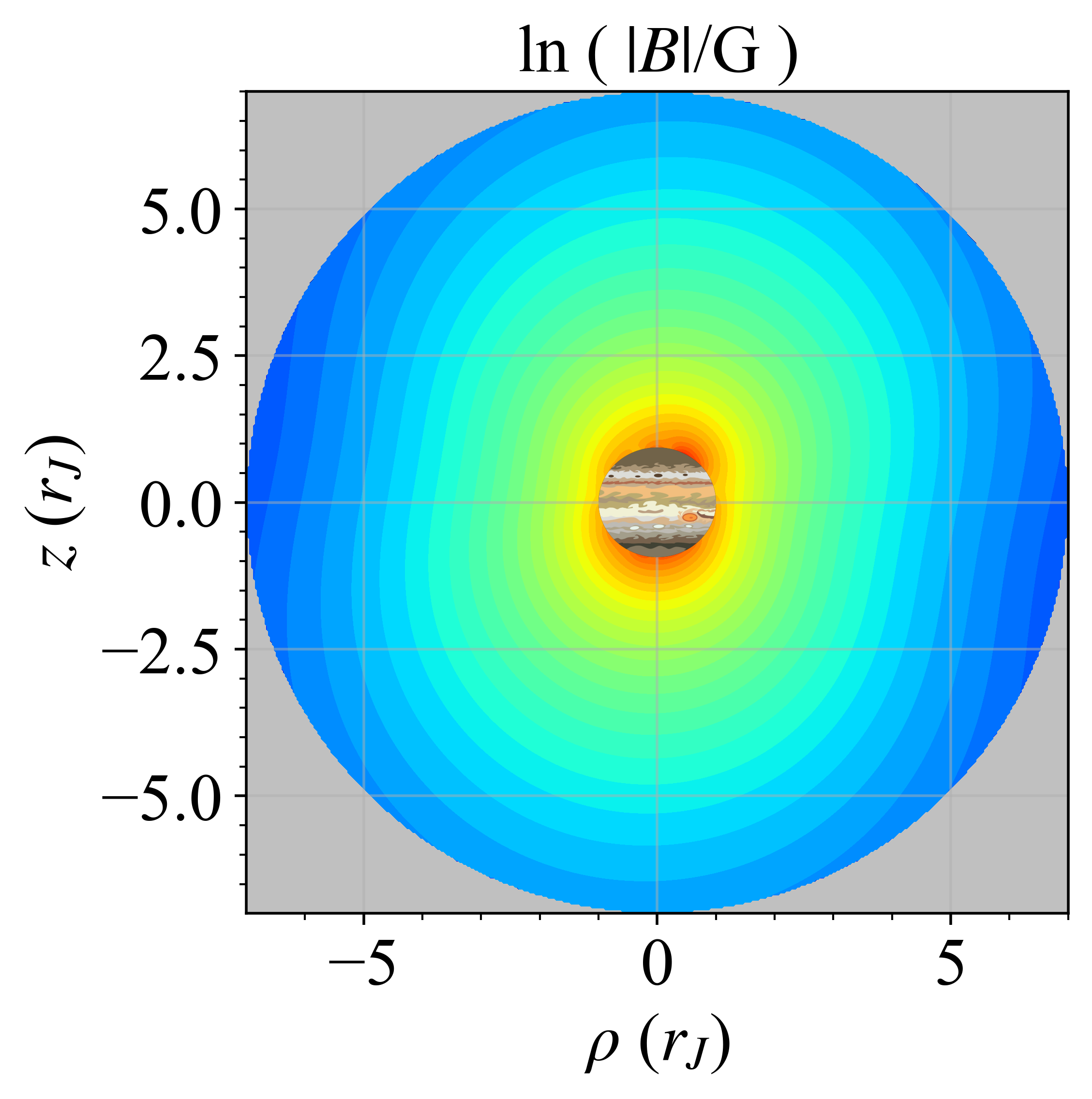}}
	\subfloat{\includegraphics[width = 0.45\textwidth]{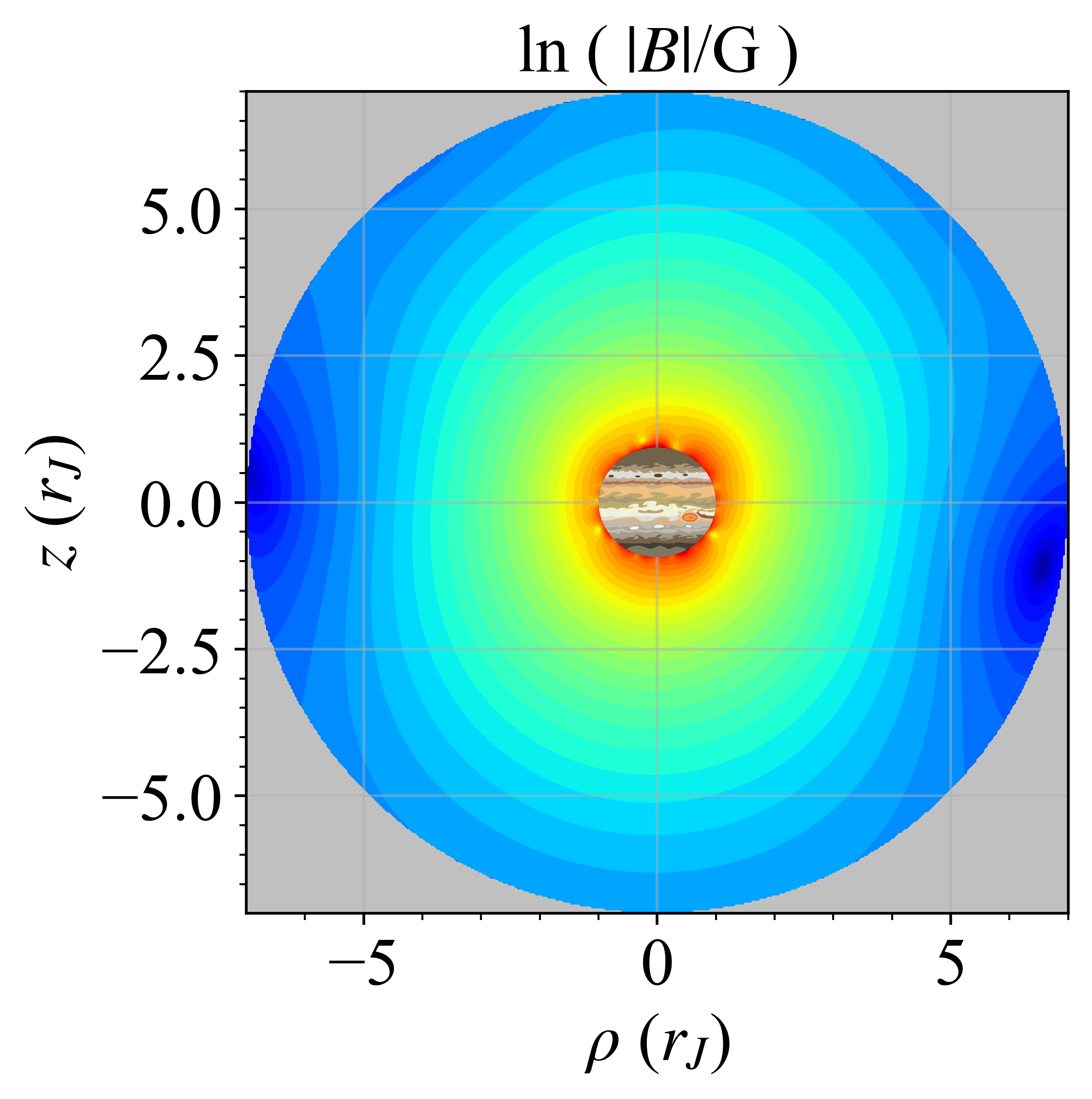}}
	\subfloat{\includegraphics[width = 0.1\textwidth]{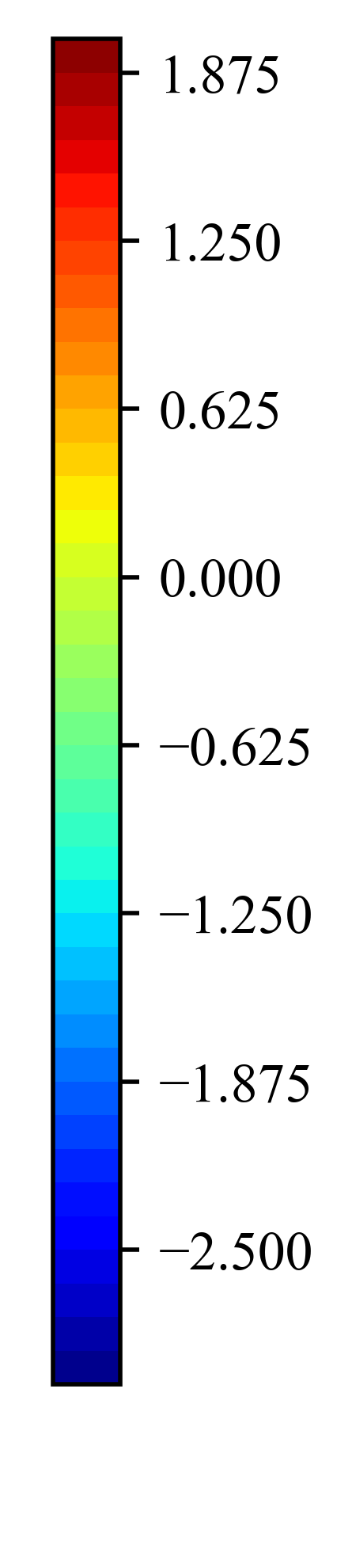}}\\
	\caption{Contours of logarithmic of the planar magnetic field (in unit of G) when $m_X = 10 ^{-13}$~eV (left) and $3 \times 10 ^{-15}$~eV (right) respectively, and $\varepsilon = 1$, on a plane between the longitudes $150^\circ$ and $330^\circ$. Here, $z$ and $\rho$ are the height and the radial distance in cylindrical coordinate system defined using the Jovian system III. The field is limited within $r = 7 r_J$ and the grey area is where the plasma currents exist. In the left panel, since $m_X$ is large, the modification to the magnetic field decays rapidly outside Jupiter. In this case, $\Delta \chi^2 = 1.07 \times 10^{-9}$ and the field configuration is almost identical to that of the SM case.}
    \label{compare_extreme_chi2}
\end{figure}

The field intensity at a particular longitudinal plane with either small (left panel) or large (right panel) $\chi^2_{\rm min}(\pmb{\psi})$ is shown in \Fig{compare_extreme_chi2}. As discussed in \Sec{math}, when dark photon mass is close to $r_J^{-1}\approx 2.8 \times 10^{-15}$ eV, its effect is maximized and the corresponding magnetic field could be modified significantly, as shown in the right panel of \Fig{compare_extreme_chi2}. The multiple anomalous points with low $|\pmb{B}|$ near the surface indicate significant non-dipole behavior, as does the high field intensity when $r \lsim 2 r_J$. This is an extreme case that is excluded, as we will discuss more in the following section. 

We then scan the $\Delta \chi^2 (\pmb{\psi}) \equiv \chi^2_{\rm min}(\pmb{\psi}) - \chi^2_{\rm min}(\rm SM)$ for both the massive photon and dark photon models, where $\chi^2_{\rm min}(\rm SM)$ is the minimum $\chi^2$ for the SM with only one massless photon. As mentioned, $\Delta \chi^2(\pmb{\psi})$ can serve as a rough estimator of the marginalized likelihood ratios between the SM and the new physics scenario. We find that the minimum value of $\Delta\chi^2(\pmb{\psi})$ is always positive, and therefore the Juno data does not show any evidence for beyond SM physics, and the result could be interpreted as constraints on new physics parameters. For the SM, we find that $\chi^2_{\min}({\rm SM})/(3 d) = 1.36$.\footnote{The JRM33 model~\cite{connerney2018new} corresponds to $\chi^2_{\min}({\rm SM})/(3 d) = 7.36$. This is due to a different method they use, which minimizes the residue of the magnetic field instead of $\chi^2$.}

\begin{figure}[h!]
    \centering
	\subfloat{\includegraphics[width=\textwidth]{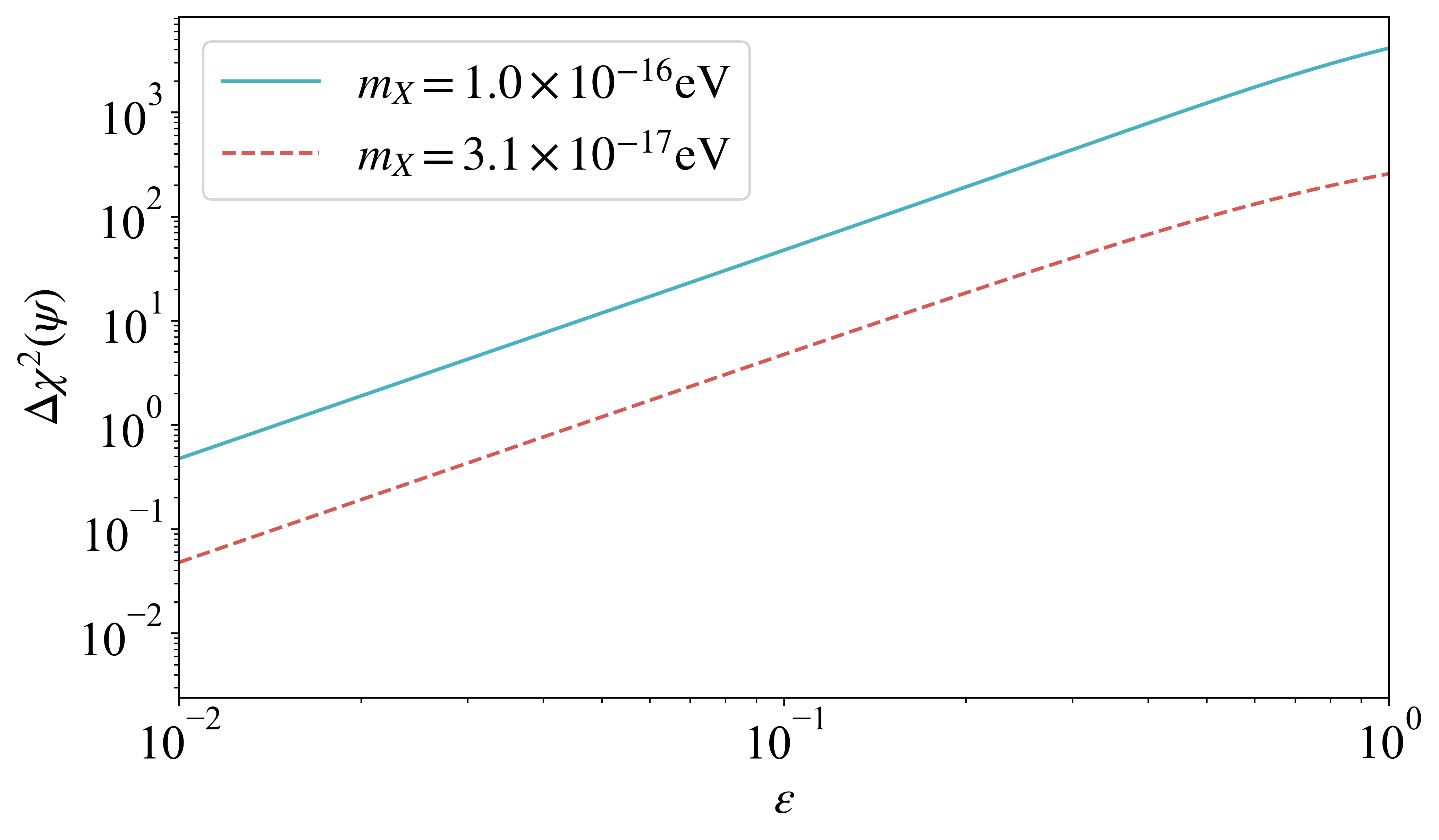}}\\
	\subfloat{\includegraphics[width=\textwidth]{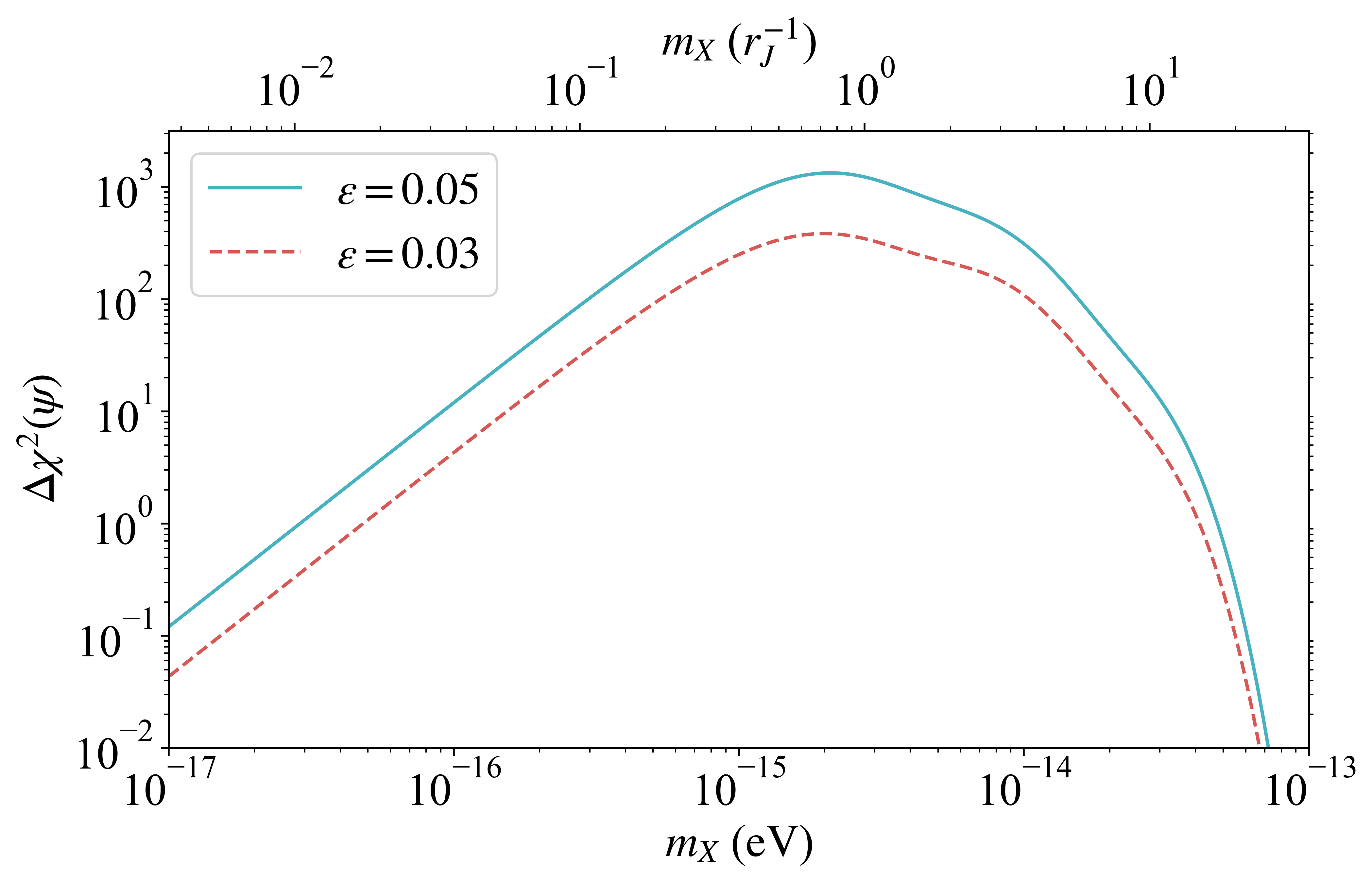}}
    \caption{Upper: $\Delta \chi^2$ as a function of $\varepsilon$ when $m_X$ is fixed at $10^{-16}$~eV and $3.1 \times 10^{-17}$~eV. Lower: $\Delta \chi^2$ as a function of $m_X$ when the $\varepsilon$ is fixed as 0.03 and 0.05.}
    \label{scan_demo}
\end{figure}

We show $\Delta \chi^2$ as a function of $\varepsilon$ for a given $m_X$ in the upper panel of~\Fig{scan_demo}. From it, one could see that $\Delta \chi^2$ keeps increasing as $\varepsilon$ increases. In the lower panel of \Fig{scan_demo}, we show $\Delta \chi^2$ as a function of $m_X$ for a given $\varepsilon$. In this case, $\chi^2$ first increases and then decreases, with a peak at $m_X \approx 2.0 \times 10^{-15}~\eV \approx 0.7~r_J^{-1}$. The minimum value of $\chi^2/(3d)$ is achieved at both low and high mass ends, which is consistent with our expectations discussed at the end of \Sec{math}, {\it i.e.}, when $m_X$ is very large or small, the model is similar to the SM, where the minimum $\chi^2$ is achieved. 

\section{New Physics Constraints}
\label{result}

In the previous section, we successfully reconstruct the Jovian magnetic field by minimizing $\chi^2$, and the result agrees with the most updated model in the geophysics literature~\cite{connerney2018new}. In this case, the parameter set $\pmb{c}$ is chosen to maximize the likelihood function $p(\pmb{B}|\pmb{\eta}$)=$p(\pmb{B}|\{\pmb{\psi} ,\pmb{c}\})$ as they are the target parameters for magnetic field modeling. Conversely, to probe new physics such as a non-zero SM photon mass or a dark photon, only $\pmb{\psi}$ is relevant. The magnetic field model parameters $\pmb{c}$ become nuisance ones as we have no direct measurement of the dynamo structure inside Jupiter that generates the $B$ field. In general, the Jovian magnetic field could be anything similar to the current configuration. Thus $\pmb{c}$ should be marginalized properly, and the likelihood function evaluates the performance of each new physics scenario by averaging over $\pmb{c}$
\begin{equation}
    p(\pmb{B}|\pmb{\psi})=\int d \pmb{c}~p(\pmb{B}|\pmb{\eta})~.
\end{equation}
In the equation above, we omit a non-trivial prior distribution of $\pmb{c}$ for two practical reasons: 1) the reconstructed magnetic field can only match the Juno measurement when it is very close to the best fit, and any non-trivial prior shall become essentially flat within such a narrow range; 2) since our likelihood function is Gaussian with $\pmb{c}$ appearing quadratically in its exponent, the marginalization is no more than a Gaussian integral. As mentioned in the previous section, the likelihood $p(\pmb{B}|\pmb{\psi})$ after marginalization has the same behaviour as the maximized likelihood $\propto e^{-\chi_{\rm min}^2(\pmb{\psi})}$.

A Bayesian analysis is then possible once a prior on the new physics parameter set $\pmb{\psi}$ is introduced. In particular, the posterior distribution $p(\pmb{\psi}|\pmb{B}^\ob)$ from the Juno MAG measurements $\pmb{B}^\ob$ reads
\begin{equation}
    p(\pmb{\psi}|\pmb{B}^\ob)=\frac{1}{N} \int p (\pmb{B}^\ob|\pmb{\psi}) \pi(\pmb{\psi})~, 
\end{equation}
where $\pi(\pmb{\psi})$ is the prior of new physics parameter $\pmb{\psi}$ and $N$ is the normalization factor. Here the choice of prior $\pi(\pmb{\psi})$ can be crucial since the marginalized likelihood approaches its local maximum at the SM-like limit, $i.e.$ when $\varepsilon$ is small, or $m_{X}$ is far away from $r_J^{-1}$. Notably, these conditions are met on the boundary of the parameter space of interest: by exploring the new data set, we will push the limit towards these SM-like regions. Therefore, the posterior distribution $ p(\pmb{\psi}|\pmb{B}^{\rm ob})$ will also reach the maximum value near the boundary if the prior is flat, making the normalization factor strongly dependent on the volume of the prior. The issue of the prior volume effect can be avoided if a reasonable prior is chosen so that $\pi(\pmb{\psi})\simeq 0$ when $\varepsilon=0$, $m_{X}\gg r_J^{-1}$, or $m_{X}\ll r_J^{-1}$. Thus we use the Jeffreys prior~\cite{1946RSPSA.186..453J}, which converges to zero near the SM-like boundaries of $\pmb{\psi}$. The Jeffery's prior of $\pi_{\rm J}(\pmb{\psi})$ is defined as
\begin{equation}
    \pi_{\rm J}(\pmb{\psi}) \equiv \sqrt{\det [\mathcal{I}(\pmb{\psi})]}~,
\end{equation}
where the matrix element of the Fisher information matrix $\mathcal{I}(\pmb{\psi})$ reads
\begin{equation}
    \mathcal{I}(\pmb{\psi})_{ij}\equiv - \int d \pmb{B}^\prime \frac{\partial^2}{\partial \pmb{\psi}_i \partial \pmb{\psi}_j} \log p(\pmb{B}^\prime|\pmb{\psi})~.
    \label{Fisher information matrix}
\end{equation}

In the equation above, we take the expectation value of $\frac{\partial^2}{\partial \pmb{\psi}_i \partial \pmb{\psi}_j} \log p(\pmb{B}^\prime|\pmb{\psi})$ by averaging it under all possible magnetic field measurements $\pmb{B}^\prime$. Therefore, the prior $\pi_{\rm J}(\pmb{\psi})$ is not determined by the data $\pmb{B}$ itself, but by its structure, $i.e.$, when and where the measurements were performed. Since the data is unable to tell the SM from new physics scenarios when $\pmb{\psi}$ lives in the SM-like limit, generically the $\pmb{\psi}$ dependence of the marginalized likelihood $p(\pmb{B}^\prime|\pmb{\psi})$ vanishes, resulting in $\pi_{\rm J}(\pmb{\psi})\simeq 0$. Consequently, the prior volume effect is alleviated, and the result becomes insensitive to the boundaries of the integration. It is also known that the Jeffreys prior leaves the result of the analysis invariant under redefinition of $\pmb{\psi}$. The approach is widely used in cosmology, astrophysics, and particle physics studies (see~\cite{Dumont:2012ee,Centers:2019dyn,Hsueh:2019ynk,Bertone:2019vsk,Fedderke:2021aqo,Fedderke:2021rrm} for examples).

Thanks to the analytical expressions of the radial functions in~\Eq{mass_R_ex} and ~\Eq{eq:darkphotonradialinternal}, the calculation of the prior $\pi_{\rm J}$ is straightforward in both the massive photon and dark photon cases.

\noindent\textbf{Photon mass limits} From the Juno MAG data, we first obtain a 95\% credible upper limit of photon mass:
\begin{equation}
    m_\gamma < 2.5 \times 10^{-18}~\text{eV}~.
\end{equation}
Compared with the well-cited limit using the Pioneer 10 data, $m_\gamma < 6\times 10^{-16}$~eV~\cite{davis1975limit}, the Juno measurement improves this limit by two orders of magnitude thanks to the significantly improved coverage and precision.

\begin{figure}[h!]
    \centering
    \includegraphics[width=\textwidth]{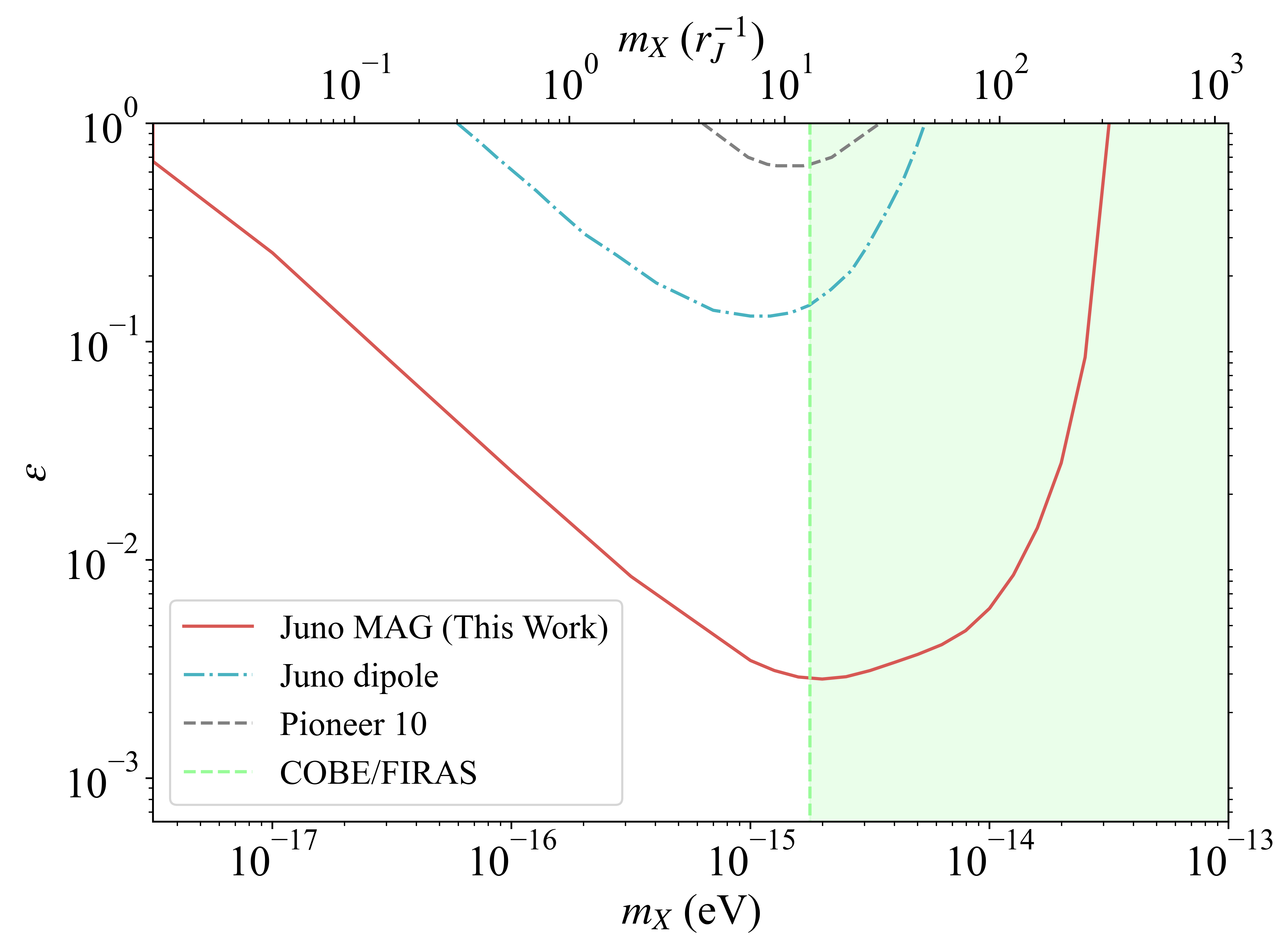}
    \caption{Constraints on the dark photon mass $m_X$ and kinetic mixing parameter $\varepsilon$ with the SM photon. The gray dashed line is the constraint using the Pioneer 10 data set; the blue dash-dotted line is from \cite{marocco2021dark}; the region to the right of the green dashed line is excluded using the data from COBE/FIRAS \cite{caputo2020dark, garcia2020effective}; and the red line is the upper bound derived using the Juno data set in this work.  }
    \label{limit_95per_1to7}
\end{figure}

\noindent\textbf{Dark photon limits} For the dark photon scenario, there are two parameters ($m_X,\epsilon$) to be probed. However, the Jeffreys prior is better suited for a one-dimensional subspace of $\pmb{\psi}$ with the rest of $\pmb{\psi}$ being fixed~\cite{2009arXiv0904.0156B}. It behaves as an objective prior that maximizes the information gained from the posterior distribution. Thus, we choose to fix $m_X$ and the Jeffreys prior $\pi(\varepsilon)$ depends on the kinetic mixing parameter only. 
To regulate the posterior distribution, the Jeffreys prior $\pi_{\rm J}(\varepsilon)$ is set to zero when $\varepsilon>1$. The red contour shows the 95\% credible limit from our analysis in Fig.~\ref{limit_95per_1to7}, overlaid with early results using the Jupiter mission data. 
At $\varepsilon = 1$, $m_X \leq 3.2 \times 10^{-18}$~eV or $m_X \geq 3.2 \times 10^{-14}$~eV. The strongest bound on $\varepsilon$, $\varepsilon \leq 2.8 \times 10^{-3}$, is achieved when $m_X = 2.0 \times 10^{-15}$~eV. Similar to the case of constraining the photon mass, the result from Juno MAG data greatly exceeds the limit using the Pioneer 10 data. Compared with the recent work, which gives the strongest constraint $\varepsilon \leq 0.13$ at $m_X = 1.1 \times 10^{-15}$~eV~\cite{marocco2021dark}, our work results in a much stronger constraint using the Juno's data. 

As shown in \Fig{limit_95per_1to7}, for $m_X \geq 1.8 \times 10^{-15}$~eV, a better constraint is acquired using the measurement from COBE/FIRAS~\cite{caputo2020dark, garcia2020effective}. This arises from the conversion of the CMB photons to dark photons that causes both spectral distortions and additional anisotropies in the CMB. On the other hand, our constraint for $3.2 \times 10^{-18}$ eV $\leq m_X \leq 1.8 \times 10^{-15}$~eV is the strongest one among all possible constraints independent of the dark photon relic density. In other words, our constraint still applies even when the dark photon is not (part of) dark matter. 

\section{Conclusions and Outlook}
\label{conclusion}

Our study underscores the significant potential of planetary science and geophysics data in constraining new physics beyond the SM. The application of static planetary magnetic field measurements serves as a classic method to place constraints on a non-zero SM photon mass and properties of light dark photon. The power of this method strongly relies on the coverage and precision of the $in \; situ$ survey of a planet's magnetic field. In this work, we focus on the Juno mission orbiting Jupiter, which hosts the largest stable planetary magnetic field in the solar system. With the Juno mission's open MAG data, we are able to set limits on the photon mass and dark photon parameters orders of magnitude more stringent than those derived from the old Pioneer missions, which are still widely quoted in the literature. This leap in sensitivity can primarily be attributed to Juno's closer proximity to the Jovian surface during its orbiting. Such close encounters enhance the sensitivity to new physics modifying the magnetic field. Additionally, the comprehensive angular coverage achieved by the Juno orbiter's 39 perijove trajectories allows for a more detailed examination of Jupiter's magnetic field, including its higher-multipole components. 

Our methodology, in the spirit of multipole expansion used in the geophysics studies, effectively handles the complicated Jovian magnetic field. However, unlike traditional geophysical studies, we adopt an updated statistical approach incorporating uncertainties in a more robust way. We identify major contributions to the variance, such as the short-term fluctuations in data and the long-term drifting of the magnetic field. Then applying our statistical approach, we reconstruct the Jovian magnetic field with a high degree of accuracy, aligning closely with the latest models widely accepted in the geophysics community. This agreement not only validates the precision of our reconstruction process but also establishes a solid foundation for further new physics exploration. 

Implementing a Bayesian analysis with the Jefferys prior in the new physics parameter space, we obtain constraints which significantly surpass those from the previous Pioneer data. In particular, for the photon mass, our method excludes a photon mass greater than $2.5\times 10^{-18}$~eV. Similarly, the constraints on light dark photon are also strong. For example, the mixing parameter $\varepsilon$ is constrained to be $ <2.8 \times 10^{-3}$ when the dark photon mass $m_X $ is $2.0 \times 10^{-15}$~eV. While our photon mass limit is close to the current best limit set by the solar wind~\cite{Ryutov:2007zz}, our method leads to the leading non-dark matter dark photon constraint when the dark photon mass is between $10^{-17}-10^{-15}$~eV.

The complexity of the Juno MAG data and Jupiter's planetary system suggests that improving the limits set by our current method would require a substantial overhaul of our analysis framework. At higher precision levels, Jupiter's magnetic field ceases to be a simple, slowly-changing field described by multiple time-independent coefficients. Instead, dynamic phenomena, such as interactions between the plasma medium in the radiation belt and external solar fields or solar wind, become increasingly relevant. In our analysis, simplifications are made to absorb the residual impacts of complex time-dependent plasma or external solar field effects within the data-driven variance term. In addition, our expansion of external fields assumes negligible plasma magnetodisk currents within $r< 7 r_J$. While supported by previous analyses, this approximation may not hold at higher precision levels. These effects, while currently subdominant at the precision level of our analysis, will inevitably need to be factored into future analyses. 

In future efforts, several advancements could further refine our limits. Another prominent question is what else can be done using the planetary science data to investigate new physics. A non-exhaustive to-do list includes:
\begin{itemize}
    \item Develop models to account for the plasma disk current, which is partly driven by Jupiter's rotation and is time-dependent. A larger set of data taken at $r\gtrsim \mathcal{O}(10) r_J$ is necessary to achieve this. Incorporating other available $in \; situ$ Juno data, like ion phase space distribution or UV spectra, may provide deeper insights. This would also be beneficial for constraining local plasma density, among other parameters, although it requires a more detailed knowledge of the magnetosphere.
    \item Aside from the SM photon mass and light dark photon scenario, a few other new physics can also modify electromagnetic fields at macroscopic scales. For instance, theories with light (pseudo-)scalars that interacts with the photon may also alter the Jovian magnetic field. A systematic survey of the Juno MAG limits on these models could be interesting. 
    \item Looking beyond the Juno mission, the next generation of exploratory missions to Jupiter's magnetosphere~\cite{grasset2013jupiter,2023SSRv..219...48K} will bring a wealth of data and help bridge the gap in Juno measurements. They may provide exciting opportunities to understand Jupiter's complex magnetic environment better and further benefit particle physics and cosmology. 
\end{itemize}

\section*{Acknowledgement}
We thank Jack Connerney, Itamar Allali, Carlos Blanco, Xucheng Gan, Rebecca Leane, Giacomo Marocco, and Linda Xu for useful comments and feedback. We also thank all the useful feedback when SY was delivering a PACMAN series talk at CCA, New York University. JF and LL thank the physics department of Harvard University for hospitality when this work is conducted. JF and LL are supported by the NASA grant 80NSSC22K081 and the DOE grant DE-SC-0010010.   

\appendix


\section{Spectrum of the Multipoles}
\label{app:spectrum}

\begin{figure}[h!]
    \centering
    \includegraphics[width=\textwidth]{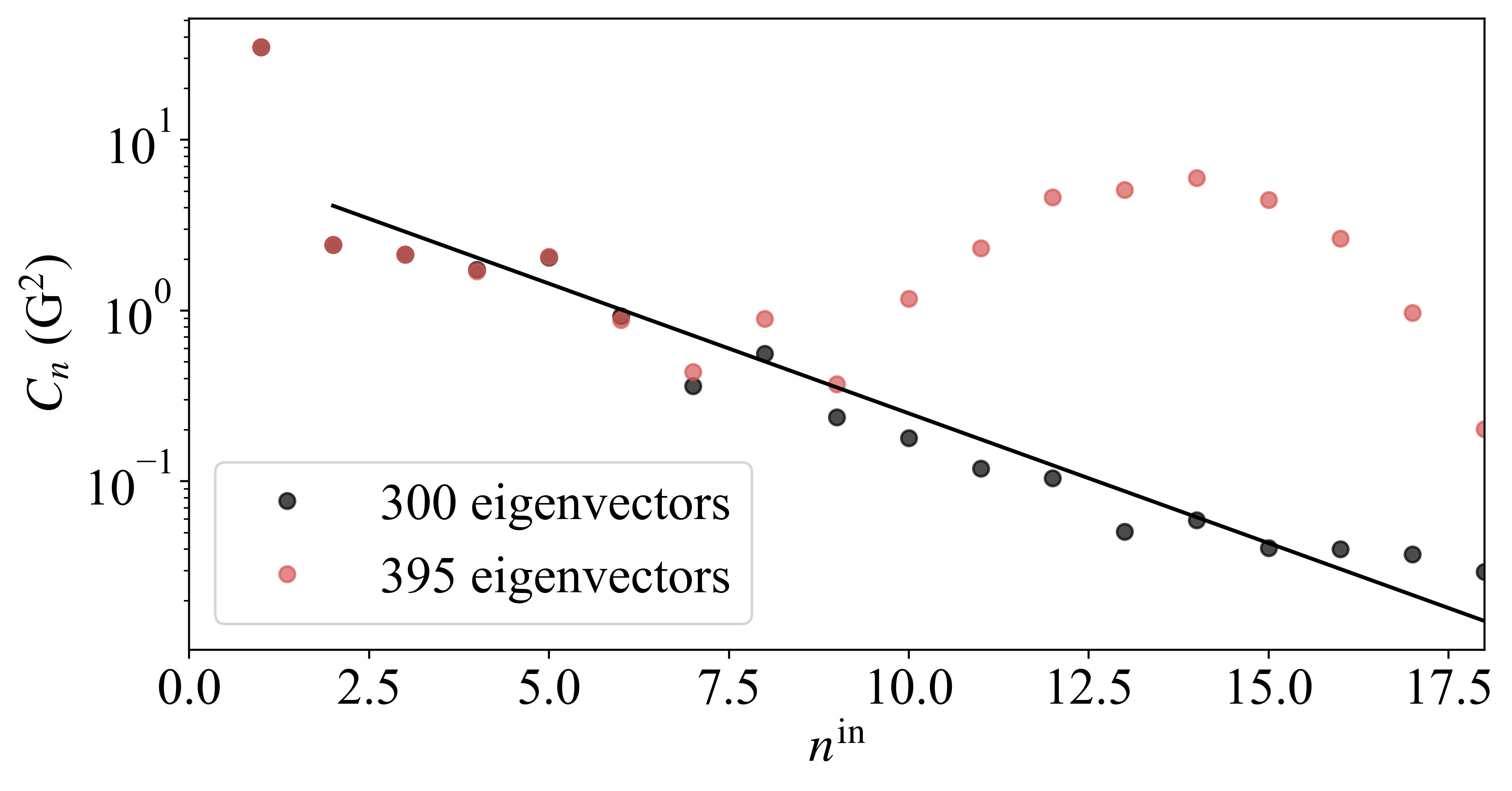}
    \caption{The magnetic spectrum for a fit to the Juno observations. The red dots represent the result with all 395 eigenvectors, which  suffer from an overfitting problem, compared with the result using only a subset of 300 eigenvectors (shown by black dots). The spectrum between $n = 2$ and $n = 18$ can be well fitted by a power law relation in $n^{\inter}$.}
    \label{spec}
\end{figure}

To determine the appropriate $n_{\max}^{\inter}$ and the number of eigenvectors of $\pmb{c}$ to use, we plot the spectrum of different multipoles with the definition given in~\cite{lowes1974spatial}
\begin{equation}
    C_n = (n + 1) \sum_{l = 0}^n [(g_n^l)^2 + (h_n^l)^2]~.
\end{equation}
The result of the fit with a massless SM photon is shown in \Fig{spec} as red dots. After filtering out the dominant dipole and quadrupole contributions, a power law relationship between $C_n$ and $n$ is expected, which is consistent with a finite core size of Jupiter~\cite{lowes1974spatial}. From \Fig{spec}, the power law relationship holds for a small $n^{\rm in}$ but is violated beyond $n^{\rm in}=9$ due to data overfitting. In particular, the sparsity of the Juno MAG measurements near Jupiter's south pole leave the relevant eigenvectors poorly constrained.
For magnetic field associated with new physics parameters close to the limit given in~\Sec{result}, the pattern would remain similar as the deviation from the SM is small.

Truncating the number of eigenvectors will significantly alleviate the overfitting problem. By removing eigenvectors with small eigenvalues and keeping only 300 of them, the deviation of the power law relation is postponed until $n^{\rm in} = 18$, as shown in \Fig{spec} as black dots. This approach is also adopted in the latest Jovian magnetic field model~\cite{connerney2022new}. We thus choose $n_{\max}^{\rm in} = 18$ and keep 300 eigenvectors as in~\cite{connerney2022new}. As a final remark, such a truncation on eigenvectors is helpful to get a proper magnetic field model. It has no significant effect in the $\pmb{\psi}$ limit since these unconstrained eigenvectors have very small eigenvalues and only change the likelihood minimally.

\bibliography{Ref}

\end{document}